\documentclass[prd,aps,pdftex]{revtex4-1} 

\usepackage{amsmath}
\usepackage{tipa}
\usepackage{amssymb}
\usepackage{graphicx}
\usepackage{xfrac}
\usepackage{bm}
\usepackage{enumerate}

\pdfoutput=1

\newcommand{\be}{\begin{equation}}
\newcommand{\ee}{\end{equation}}
\newcommand{\bea}{\begin{eqnarray}}
\newcommand{\eea}{\end{eqnarray}}
\newcommand{\ba}{\begin{eqnarray}}
\newcommand{\ea}{\end{eqnarray}}
\newcommand{\GeV}{\,\mathrm{GeV}}

\begin{document}

\title{ The sounds of the Little and Big Bangs}


\author{Edward Shuryak}


\affiliation{%
Stony Brook University, edward.shuryak@stonybrook.edu}




\begin{abstract}

Studies of heavy ion collisions have discovered that tiny fireballs of new phase
of matter -- quark gluon plasma (QGP) -- undergoes explosion, called the Little Bang.
In spite of its small size, it is not only well described by hydrodynamics, but even small
perturbations on top of the explosion turned to be well described by hydrodynamical sound modes.
The cosmological Big Bang also went through phase transitions, the QCD and electroweak
ones, which are expected to produce sounds as well. We discuss their subsequent 
evolution and  hypothetical inverse acoustic cascade, amplifying the
amplitude. Ultimately, collision of two sound waves leads to formation of gravity waves,
with the smallest wavelength. We briefly discuss ho those can be detected. 

\end{abstract}

\maketitle










\setcounter{section}{-1} 

\section{Introduction}
\subsection{An outline}
   This paper is a short review describing some
   recent developments in two very different fields, united by some common physics
   but being at very different stages of their develoment.

   One of them is Heavy Ion Collisions, creating the {\em Little Bangs}
  mentioned in the title. In an explosion, lasting in a small volume 
  for very short time, they recreate the hottest matter created in the laboratory
   -- known as Quark-Gluon Plasma (QGP)-- which was not present 
   in Universe since the early stages of its cosmological evolution, the {\em Big Bang}.
 QGP turned out to be rather unusual fluid, and we will briefly discuss 
 why we believe it is the case. But the common topic which holds two parts of this paper together are 
   {\em the sounds}, mentioned in the title. As always, those are  small 
   amplitude perturbations of hydrodynamical nature.   Not unusual by themselves,
   they still surprised us, since nobody expected them  to be experimentally
   detected in a system as small as the Little Bangs. The latest developments has shown
   that in fact they are to certain extent present even is smaller systems, such as central proton-nuclei
   collisions and even in proton-proton collisions, in events with unusually high multiplcity.
   
   In cosmological settings sound perturbations have been rather well studied, using perturbations
   of the Cosmic Microwave Background (CMB), and, to some extent, correlation functions
   of Galaxies. However, 
   those correspond to relatively late stage of the Universe, at which the temperature is low enough for matter
   to get de-ionized, with the temperature $T<1 \, eV$.
   We will touch these phenomena only peripherally, because
   of certain similarities to sounds in the Little Bang. 
   The  main question we will be discussing in the second part of the paper is the title of section \ref{sec_cosmo}:
   Are cosmological phase transitions observable? 
   Transitions are in plural because we mean here both the {\em electroweak}
   transition, at $T_c\sim 100\, TeV$
    and
the {\em   QCD phase transitions} at  $T_c\sim 160\, MeV$. We hope the answer is affirmative, but one still has to figure out how it can be done.
   
   A specific scenario we will discuss is a possibility of {\em inverse acoustic cascade},
   which can carry sounds, from the UV end of the spectra, with momenta $p\sim \pi T\sim 1\, GeV$,
   (for QCD) and $\sim 1 \, TeV$ for electroweak transition, to the IR end of the spectra
   provided by the cosmological horizon at the corresponding times. If such cascade
   is there, it works like an powerful acoustic amplifier. At the end of the process,
   two sound waves can be converted  into a  gravity wave, 
   which survive all the later eras and can be potentially detectable today.
   
   \section{The Little Bang}
   \subsection{The quest for Quark-Gluon Plasma}
  Aiming at  non-experts, we start with motivations and brief history of the 
  field. What was the reasons to study high energy Heavy Ion Collisions?
  What has been found, and why it is rather different from what is
  observed in high energy $pp$ collisions?

    There are three
   different (but of course interrelated) aspects of it. One is the $theoretical$ path, coming from
   1970's after the discovery of QCD, first in its perturbative form, and then in
   a non-perturbative theory. Development of QCD at finite temperature and/or density
   lead to realization that QGP is a completely new phases of matter.
   Now work in this direction includes not only certain
   number of theorists, specializing in QFTs and statistical physics, but also a community performing
   large scale computer simulations of lattice gauge theories, and rather sophisticated models based on them. This activity also has grown up and includes collaborations of 
   dozens of people. As we will discuss below, QGP is a very peculiar plasma,
   with rather unusual kinetic properties. We will discuss one proposed explanation of that,
based on the fact that this plasma  includes both electric and magnetic charges, 
   
   The second (and now perhaps the dominant)  aspect of the quest for QGP is the $experimental$ one.    Let me here just mention that
   experimental activity is now dominated by five large collaborations:
   STAR and PHENIX (now under complete rebuilding of the detector) at Relativistic heavy ion collider (RHIC) at Brookhaven National Laboratory,  and ALICE,CMS and ATLAS at 
   Large Hadron Collider
   (LHC) at CERN. The last two have been basically built by high energy physics community and designed for
   other purposes, but both also work just fine for heavy ion collisions as well,  recording thousands of secondaries per event. 
     Each of the collaborations have hundreds of members, so
the    ``Quark Matter" and other conferences on the subject has become huge in size, and
obviously dominated by experimental talks. It is completely justifiable, as the 
list of discoveries -- often puzzling or at least unexpected -- continues.

   We will only focus on data indicating collective flows of QGP, including
   its perturbations in connection with the sound waves. 
   Of course, there are many different aspects of heavy ion collisions which we will not
   touch upon in this short text. In particular, we will not discuss dynamics of jet quenching, of heavy flavor quarks/hadrons, large event-by-event fluctuations perhaps indicative to QCD critical point, etc. For a  more complete recent review, aimed at experts, see \cite{myRMP}.

    The third direction to be discussed below is related with certain connections which
the QGP physics have  with    {\em   cosmology}. 
  Today's cosmology  is not just an intellectually challenging field, but it is now among the most rapidly developing   parts of physics. And yet, since  QGP/electroweak plasma in the early Universe happened 
  at rather early stage,  it remains challenging to find any observable
  trace of its presence.  
  It is even more so for the electroweak plasma, undergoing a phase transition into a 
  ``Higgsed" phase we now live in.
  So, very few people think about, and even those do, turn to it intermittently. 
  
Covering brief history of the QGP physics, let me 
   follow a time-honored tradition of the historians and divide it into {\em three periods}
   called (i) pre-RHIC, (ii) the RHIC era, and (iii) RHIC+LHC era. 
   
   The first period was the longest one, it started at mid-1970's and lasted for a quarter of a
   century, till the year 2000. While there were important experiments addressing heavy ion collisions in fixed target mode, at CERN SPS and Brookhaven AGS accelerators, 
   it is fair to say that in this period  the experimental program and the whole community only
   started to be built. Most talks at the conferences of that era were 
   theory-driven. 
   
   Since the start of the RHIC era in 2000, it has become soon apparent
that the data on particle spectra show evidences to strong collective flows. 
Those,  especially the quadrupole or $elliptic$ flow, confirm nicely predictions of hydrodynamics.
Most successful were hydro codes
supplemented by hadronic cascades at freezeout \cite{Teaney:2000cw,Teaney:2001av,Hirano:2005xf},
as they correctly take care of the final (near-freezeout) stage of the collisions.
All relevant dependences -- as a function of $p_\perp$, centrality, particle mass, rapidity and collision energy -- were
checked and found to be in good agreement. Since the
famous 2004 RBRC workshop in Brookhaven, with theory and experiment summaries collected in a special volume, Nucl.Phys.A750 ,  the statements that QGP "is  a near-perfect liquid"
which does flow hydrodynamically has been repeated many times since. 

The theorists at this point had recognized  that QGP in these conditions should be  in the special, {\em strongly coupled regime}, now called sQGP for short
, and hundreds of theoretical  papers have been written, developing gauge field dynamics at strong coupling.  It was a a very fortunate coincidence, that at the same time (from mid-1990's) string theory community invented a
wonderful theoretical tool, the
AdS/CFT duality, connecting strongly coupled gauge theories to 5-dimensional
weakly coupled variants of supergravity. We will not be able to discuss this direction,
as it needs a lot of theoretical background. Let me just mention that it shed an entirely new light
 on the process of QGP equilibration, which is dual to a process of (5-dimensional)
 black hole formation. The entropy produced in a Little Bang is nothing else but
  the information classically lost to outside observers, as some part of a system happen to be inside  the ``trapped surfaces". 
 
 We will also not go into discussion of
  other strongly coupled systems which has been
also addressed by theorists and their similarity to sQGP noticed. Those include a strongly coupled classical QED plasmas at one end , and quantum ultracold atomic gases in their ``unitary"
regime at the other. These studies focus on
unusual kinetic properties, essentially unusually small mean free paths, which such systems display.

   The last (and so far the shortest) era started in the year
   2010, when the largest instrument of high energy/nuclear physics, LHC at CERN, had joined the quest for QGP. These experiments  confirmed what has been learned at RHIC and, due to their
   highly sophisticated detectors and experience collaboration teams, has  added invaluable  
   additions to what we know about its properties. Perhaps the most surprising discovery made at LHC was that
   QGP and its explosion does not happen only in heavy ion collisions. Central
   $pA$ and even high multiplicity $pp$ had shown (in my opinion, beyond any reasonable doubt) 
   
   \subsection{Thermodynamics and  screening masses of QGP} \label{sec_T}
   Omitting the ``prehistoric" period before QCD was discovered in 1973, we start at the time when  QCD  was first applied for description of hot/dense matter.
  At high $T$  the typical momenta of quarks and gluons have  scale $T$, and, due to
  asymptotic freedom,   the coupling is expected to be small \be \alpha_s={g^2(T) \over 4\pi} \sim {1 \over log(T/\Lambda_{QCD})} \ll 1 \ee 
  so it was promptly suggested  be { Collins and Perry} \cite{Collins:1974ky}
  and others,  that the  high temperature (and or
density) matter should be close to an 
ideal gas of quarks and gluons. 

There remained however the following important question: since the asymptotic freedom means that
in QCD (unlike in QED and other simpler theories) the charge is $anti$-screened by 
virtual one-loop corrections. Will there be screening or anti-screening by  thermal quarks and gluons. 
 The calculation of the polarization tensor  \cite{Shuryak:1977ut}
 have shown that unlike the virtual gluon loops which anti-screen the charge,
the real in-matter gluons behave more reasonable and $screen$ the charge: therefore
this new phase I called 
{\em Quark-gluon Plasma}, QGP for short. This happens at the so called {\em electric scale}
given by the electric screening (Debye) mass $M_E$
\be {M_E \over T} = g(T) \sqrt{1+{N_f \over 6}}, \,\,\,\,  {M_M \over T} =0 \ee 
  The second statement,  found from the same polarization tensor 
  \cite{Shuryak:1977ut}, tells  us  that in the perturbation theory static
magnetic fields are $not$ screened.
First re-summation of the so called ring diagrams
 produced a finite plasmon term 
   \cite{Shuryak:1977ut,Kapusta:1979fh}, but higher order diagrams are 
still infrared divergent. 
In general, infrared divergences and other
non-perturbative phenomena survive in the
 magnetic sector, even at very high $T$. 
 

%
%

Jumping over decades of work, let us discuss the values  of the electric and magnetic 
screening masses extracted from various approaches of today.
Those are listed in Table 1, including  predictions from various strong coupling approaches:
 the first line corresponds to a (large $N_c$) holographic model, the next two to lattice
 (the last with small physical quark masses), and the last to  the dimensionally reduced 3D effective theory for  $N_f = 3$ light quarks. Looking at this Table, one finds 
 that the electric mass is $not$ much smaller than the
 temperature: instead $M_E/T > 1$. This means the coupling is not small and pQCD is not  applicable.
 Second important observation:
  while the magnetic mass is still smaller than the electric one, it is smaller only by a factor of 2 or so.  This means magnetic charges play a significant role,
 comparable to that of its electrically charged quasiparticles, quarks and gluons.
Below we will discuss the role of magnetically charged quasiparticles, the $monopoles$,
which are believed to play an important role in QGP dynamics.

 
%
%
\begin{table}[h]
\caption{The electric and magnetic screening masses, normalized to the temperature. The last column is the square of their ratio.}
\begin{center}
\begin{tabular}{|c|c|c|c|}
\hline
reference & $M_E/T$ & $M_M/T$ & $M_E^2/M_M^2$ \\
\hline
 \cite{Bak:2007fk} & 16.05 & 7.34 & 4.80 \\
\cite{Maezawa:2010vj}  & 13.0 (11) &  5.8(2) & 5.29 \\
 \cite{Borsanyi:2015yka} & 7.31(25) & 4.48(9) & 2.66 \\
 \cite{Hart:2000ha}  & 7.9(4) & 4.5(2)  & 3.10 \\
 \hline  
\end{tabular}
\end{center}
\label{tab_screening}
\end{table}%

\begin{figure}[h]
\begin{center}
\includegraphics[width=8cm]{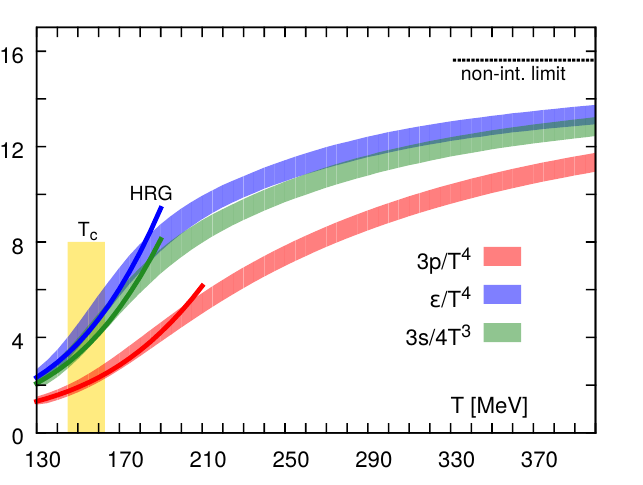}
\caption{Continuum extrapolated results for pressure, energy density and entropy.  Solid lines on the low temperature side correspond to results obtained from hadron resonance gas (HRG) model calculations.  The (yellow) band marked $T_c$ indicate the phase transition region for deconfinement and chiral symmetry restoration.}
\label{fig_thermo}
\end{center}
\end{figure}

Let us end this section with brief summary of the
{\em QCD thermodynamics on the lattice}, a numerical way to  calculate  the thermodynamical observables 
from the first principles. the QCD Lagrangian, using numerical simulations of he gauge and quark fields
discretized on a 4-dimensional lattice in Euclidean time. For a recent review see e.g.\cite{Ding:2015ona},
from which we took  Fig. \ref{fig_thermo}. 
The quantities plotted are the pressure $p$, the energy density $\epsilon$ and the entropy density $s$.

Strong but smooth rise of all quantities plotted indicate smooth but radical phase transition, from the curves marked HRG (hadron resonance gas).
The first thing to note is that quantities plotted are all normalized
to corresponding powers of the temperature given by its dimension: so at high $T$ the QGP becomes approximately scale-invariant, corresponding
to  $T$-independent constants at the r.h.s. of the plot.   The second thing to note is that these constants seem to be lower than the 
dashed line at high temperatures, corresponding to a non-interacting quark-gluon gas.
Interesting that the value for infinitely strongly interacting supersymmetric plasma is predicted to be $3/4$ of this 
non-interacting value, which is not far from the values observed.

Temperature range scanned in heavy ion experiments has been selected 
to include the QCD phase transition.
The matter produced at RHIC/LHC has the initial temperature $T\approx 2T_c$, and the final one, at the kinetic
freezeouts of the largest systems, 
is low as  $T\approx 0.5T_c$. 
While this happened more or less due to accidental factors -- like the size of the 
tunnels used for RHIC and LHC construction, and
the magnetic field in superconducting magnets available  -- 
it could not be better suited for studies of  the near-$T_c$ phenomena.

   \subsection{sQGP as the most perfect fluid} \label{sec_fluid}
   One may think, in retrospect, that development of the working model of the 
   Little Bang was a rather straightforward task: all one needed to do was to
   plug the QGP equation of state into the equations of relativistic hydrodynamics,
   and solve it with the appropriate initial conditions. 
   This was indeed so, modulo some complications. Some of them were at the freezeout stage,
   solved via switching to hadronic cascades at the hadronic phase $T<T_c$. Some of them
   were at the initial stage, such as to define the exact ``almond-shaped" fireball, created
   at the overlap of two colliding nuclei, separated by the impact parameter $\vec b$.
   
   The main difficulty on the way was  psychological:  it was completely unclear if the 
   macroscopic approach has any chance to work. 
   Most theorist were very skeptical.       Also, among firmly known facts known prior to
   RHIC experiments, was that for the ``minimally biased" 
   (typical) $pp$ collisions it did $not$ worked.
   Indeed, no collective effects --
signals of the flows -- were observed. And 
the change from one proton to nuclei 
is not numerically large, since even the heaviest nuclei are not that large.    
    
   The particular observable most watched was the so called ``elliptic flow" (see Appendix A for definition) induced by geometry of the system: at nonzero impact parameter it is an almond-shaped in the transverse plane. 
    Parton cascade models predicted that partons traveling along the longer side of 
    the almond will create more secondaries, so they predicted small and $negative$
    $v_2$. Hydrodynamics, on the other hand, 
    predicted higher pressure gradient along the shorter side of 
    the almond, and thus larger and $positive$ $v_2$.
  The very first data from RHIC decided the argument: predictions of hydro+cascade models  were confirmed, as a function of transverse momentum, centrality, particle type etc.
  The present day  
   hydro+cascade models  do it on event-by-event basis, starting from certain 
   ensemble of initial state configurations. They do excellent  job
in describing the RHIC/LHC data, see e.g.s \cite{Heinz:2013th} for review.

  
    \subsection{Sounds in the Little Bang}
     After the average pattern of the fireball explosion has been firmly established, by 2004 or so, the next goal was to study
       fluctuations, or deviations from it on event-by-event basis. 
   
   According to hydrodynamics, any small perturbation of a flow can be described
   in terms of elementary excited modes of the media. Those are longitudinal sound waves and transverse ``diffusive" mode, also associated with vorticity. So far we only have evidences for the former ones, the subject of this section.
   
   Before we proceed, let me add the following comment. An existence of sound     
 in various media is a well known fact (e.g. we use sound in air for communication), and their finding in a QGP fireball 
 may not look at first glance very exciting.   Note however, that we speak of fireballs 
 of a size of atomic nuclei, only $10 \, fm$ or so across, containing say $\sim 10^3$ particles.
 Taking a cubic root, one realizes that it is just $10\times10\times 10$ particles. Most theorists
 could not believe, prior to RHIC experiments, that such small system can show any
 collective hydro effects at all. To observe sounds $inside$ this tiny fireball
 is really a triumph, brought both by luck (a very unusual fluid, sQGP) as well as huge statistical power
      of LHC detectors. It would not be possible for any gas or drop of water, for such a small system.
      
      Let me now explain the physics of it using analogy with waves on the sea. Suppose
 somewhere near Japan there is an Earthquake, producing tsunami wave across Pacific. 
 Suppose we can only observe its consequences from very large distances,
  say from the coast
 of America. It can still be done by a correlation of small signals, like it is done 
 for now famous detection of gravity waves. 
 Say, there are two detectors, in Canada and somewhere in Chile. By correlating 
 their signals, shifted by the appropriate amount of time needed for the wave to come there,
 one  may be able to extract the correlation of sea waves and tell it  from a random noise. 
 
 This proposition may look as an unlikely scenario: but, as we will see shortly, 
 RHIC/LHC experiments do observe correlated of emission of  secondaries, 
 separated by an angle of about 120 degrees (nearly opposite sides
 of the fireball). What  one needs for that is large number of  events, to get rid  statistically
 of the random noise
. Not going into detail, consider
  few relevant numbers. Typically, there are about $10^9$ events,  each with the multiplicity $\sim 10^3$. So the number of pairs of secondaries
  is about $\sim 10^{2*3+9}$, a huge number. In fact, correlations of not just two, but also
  4 and 6 secondaries have been measured. It is enough to detect even rather weak perturbations of the fireball.

   Theoretical evaluation of these correlations proceed in 
   two stages. At first, it was done by a Green function method,
   with a delta-function like source and  linearized equation (riding on top of the average explosion,
   of course).   One group was myself and my student Pilar Staig, another lead by Ollitrault
   and the Brasilian group (Kodama, Grassi et al). It's high point was  at Annecy Quark Matter
   conference of 2011, in which these theory prediction for the shape of the correlation function and 
   the relevant data were shown, basically one after another. Their good agreement was rather shocking, even for experienced physicists.
        Then the Brasilian group pioneered the so called $event-by-event$ hydro,
 performed  for an ensemble of certain ``realistic" initial conditions. 
 This approach now became a mainstream industry, with several group
 developing it further, and finding, with
     satisfaction, that it  works spectacularly. Several angular moments of the flow perturbation 
 as a function of transverse momentum, particle type and  centrality $v_n(p_t,m,Np)$
 are reproduced.

%

 The calculations typically start  from  initial state,
 which includes geometrical shape important for elliptic flow (harmonics $n=2$) 
 and random perturbations created by particular locations of the nucleons and 
 their relative impact parameters in the collision. 
 The role of geometry reduces toward the central collisions.
 
  The dependence
 on the harmonics amplitude on their number $\epsilon(n)$ are basically independent on $n$. What that tells us is that statistically independent ``elementary perturbations" 
  have small angular size $\delta \phi\ll  2\pi$, so we basically deal with 
  a ``white noise", an angular Fourier transform of the delta function. 
 Their magnitude depends on the number of statistically independent ``cells"
 \be <\epsilon_n> \sim {1  \over \sqrt{N_{cells}} }\ee
 in the transverse plane, and this tells us what the centrality dependence of the effects should be.
Models of the initial state give us not only the r.m.s. amplitudes, but also their distribution and even correlations. 
  Remarkably, the experimentally observed  
 distributions over flow harmonics $v_n$   directly reflect those distributions
  $P(\epsilon_n)$. That means hydrodynamics does not generate any noise by itself.

      There is a qualitative difference between the main (called radial) flow and 
      other angular harmonics.
   While the former  is driven by the sign-definite outward pressure gradient, and thus
   monotonously grows with time, 
   the higher angular harmonics are
   basically sounds, and thus they behave as some
 damped oscillators. Therefore the signal observed should, on general grounds, be
  the product of the two factors: (i) the amplitude reduction due to losses, or {\em viscous damping },
 and (ii)  the  {\em phase factor} depending on the oscillation phases  $\phi_{freeze out}$,
   at the so called system freezeout.  time.
 
Let us start with the
 ``acoustic systematics"  which includes   the viscous damping factor only. It provides good qualitative account of the data
 and hydro calculations into a simple expression, reproducing  
  dependence on the viscosity value $\eta$, the size of the system $R$  and the harmonic number $n$ in question.
Let us motivate it  as follows.  
The micro scale is the particle mean free path, and 
the macro scale is the system size. Their ratio
 can be defined with the viscosity-to-entropy-density dimensionless  ratio
 \be {l \over L}= {\eta \over s}{1 \over L T} \ee 

The main effect of viscosity  on sounds is the damping of their amplitudes.
   The expression for that \cite{Staig:2010pn}  is   \be
{A(t)\over A(0)}
 = {\rm exp}\left(-{2 \over 3} {\eta \over s}
{k^2 t \over  T } \right) \label{eqn_visc_filter} \ee 
Since the scaling of the freeze out time is linear in $R$ or $t_f \sim R$, and the wave vector $k$ corresponds to
the fireball circumference which is $m$ times the wavelength
\be 2 \pi R =m {2\pi \over k} \ee 
 the expression (\ref{eqn_visc_filter}) yields
\be log({v_n\over \epsilon_n})\sim - n^2  \left({\eta \over s}\right)\left({1 \over TR}\right)  \label{eqn_acoustic}  \ee
Thus  the viscous damping is exponential of 
the product of two factors,  $\eta/s$ and $1/TR$,  each of them small, times  
 the harmonics number  squared.
    Extensive comparison of this expression with the AA data, from central to peripheral, has been
 done in Ref.  \cite{Lacey:2013is}: all its conclusions are indeed observed.
    So,  the acoustic damping provides the correct systematics
of the harmonic strength. This increases our confidence that -- in spite of somewhat different geometry --
the perturbations observed are actually just a form of a sound waves.

  For central PbPb LHC collisions with both small factors 
$\sim (1/10) $, their
  product  is $O(10^{-2})$. So one can immediately see from this expression
why  harmonics  up to  $m=O(10)$ can be observed. The highest harmonics really observed
is actually $m=6$.
Proceeding to smaller systems, by keeping a similar initial temperature $T_i\sim 400 \, MeV \sim 1/(0.5\, fm)$
but a smaller size $R$,  results in a macro-to-micro parameter that is no longer small, 
$1/TR\sim 1$, respectively. 
For a usual liquid/gas, with $\eta/s\gg 1$, there would not be any small parameter left and one would have to conclude
that hydrodynamics be inapplicable. However, since the quark-gluon plasma is an exceptionally
good fluid with a very  small  $\eta/s$, one can still observe harmonics up to  $m= 3$, even
for the small systems.


Now, if one would like to do actual hydrodynamical calculation, rather than a simple
damping evaluation by a ``pocket formula" just discussed, 
the problem appears very complicated. Indeed, the events have multiple shapes, describe by multidimensional 
probability function $P(\epsilon_2,\epsilon_3...)$. Except that it is not. All those shapes are however just a
statistical superposition of relatively simply phenomenon, a somewhat distorted analog of
an expanding circle from a stone thrown into the pond. 

Since  columns of nucleons sitting at different locations of the transverse plane    
 cannot possibly know about each other fluctuations at the collision moment, they must be statistically independent.   
   A ``hydrogen atom" of the problem is just one bump, of the size of a nucleon, on top of a smooth 
   average fireball, and all one has to do to reproduce the correlation function is to calculate 
   the Green function of the $linearized$ hydrodynamical equation.
A particular model of the initial state
expressing locality and statistical independence of ``bumps" has been formulated in \cite{Olli}: 
the correlator of fluctuations is given by the simple local expression 
\be <\delta n(x) \delta n(y)> = n(x)\delta^2(x-y)  \ee

\begin{figure}[t]
  \begin{center}
  \includegraphics[width=6cm]{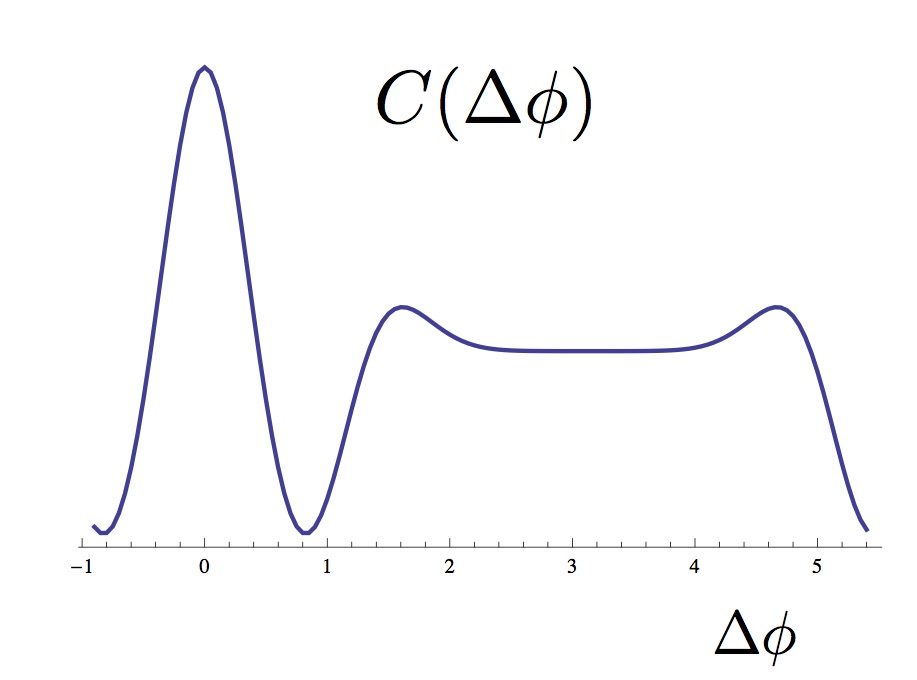}
  \includegraphics[width=6cm]{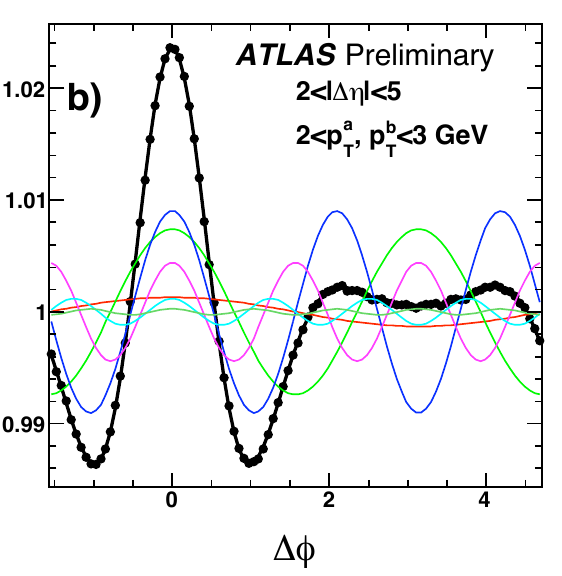} 
   \caption{(a) The two-pion distribution as
a function of azimuthal angle difference $\Delta\phi$, for  viscosity-to-entropy ratios $\eta/s=0.134$ \cite{}.
(b) from  ATLAS report \cite{ATLAS_corr}. 
 Both are   for central collisions.}
  \label{fig_2pcorr}
  \end{center}
\end{figure}

In order to calculate perturbation at later time one needs to calculate the Green function
$G(x,y)$, from the 
original location $x$  to the observation point $y$.
It has been  done 
by (my student)  P.Staig and myself \cite{Staig:2011wj} analytically, since for Gubser flow one can show that in co-moving 
coordinates all four of them can be separated. Not going into details of this excersize,
let me just note that that analytic calculation included viscosity.
 The predicted correlation function of two secondaries in central collision, as a function of relative azimuthal angle,
  is
shown in Fig.    \ref{fig_2pcorr}(a). The central feature is that there 
is one central peak, at  $\delta\phi =0$, and two more
 peaks, at $\delta\phi = \pm 2$ radian. 
 Their origin is simple and can be easily understood as soon as it is recognized
 that the main perturbation at freezeout is  located at the intercept 
 of the ``sound circle" and the fireball edge. Projected onto the transverse
 plane both are circles, of comparable size, so the  intercepts are just two points. 
 The peak at  $\delta\phi =0$ appears when both observed secondaries
 come from the same point: the radial flow thus carry them in the same direction.
 The peaks at $\delta\phi = \pm 2$ rad 
corresponding to one particle coming from one  intercept, and the other at the other. The particular angle -- about 1/3 of the circle -- appears because the sound horizon radius $R_h=c_s \tau_{freezeout}$ happens to be
numerically close to the fireball radius. As expected its area is about twice that of the other peaks.

This calculation has been presented at the first day of Annecy Quark Matter $before$  the experimental data.
The ATLAS  correlation function  (for ``super-central bin", with  the fraction of the total cross section 0-1\%)
presented a bit later is shown in Fig.    \ref{fig_2pcorr}(b). The agreement of the shape is not perfect -- because
a model is with conformal QGP and a bit different shape -- but all elements of its shape are there.

%

\subsection{Relation to the sounds of Big Bang }

\begin{figure}[t]
  \begin{center}
  \includegraphics[width=8cm]{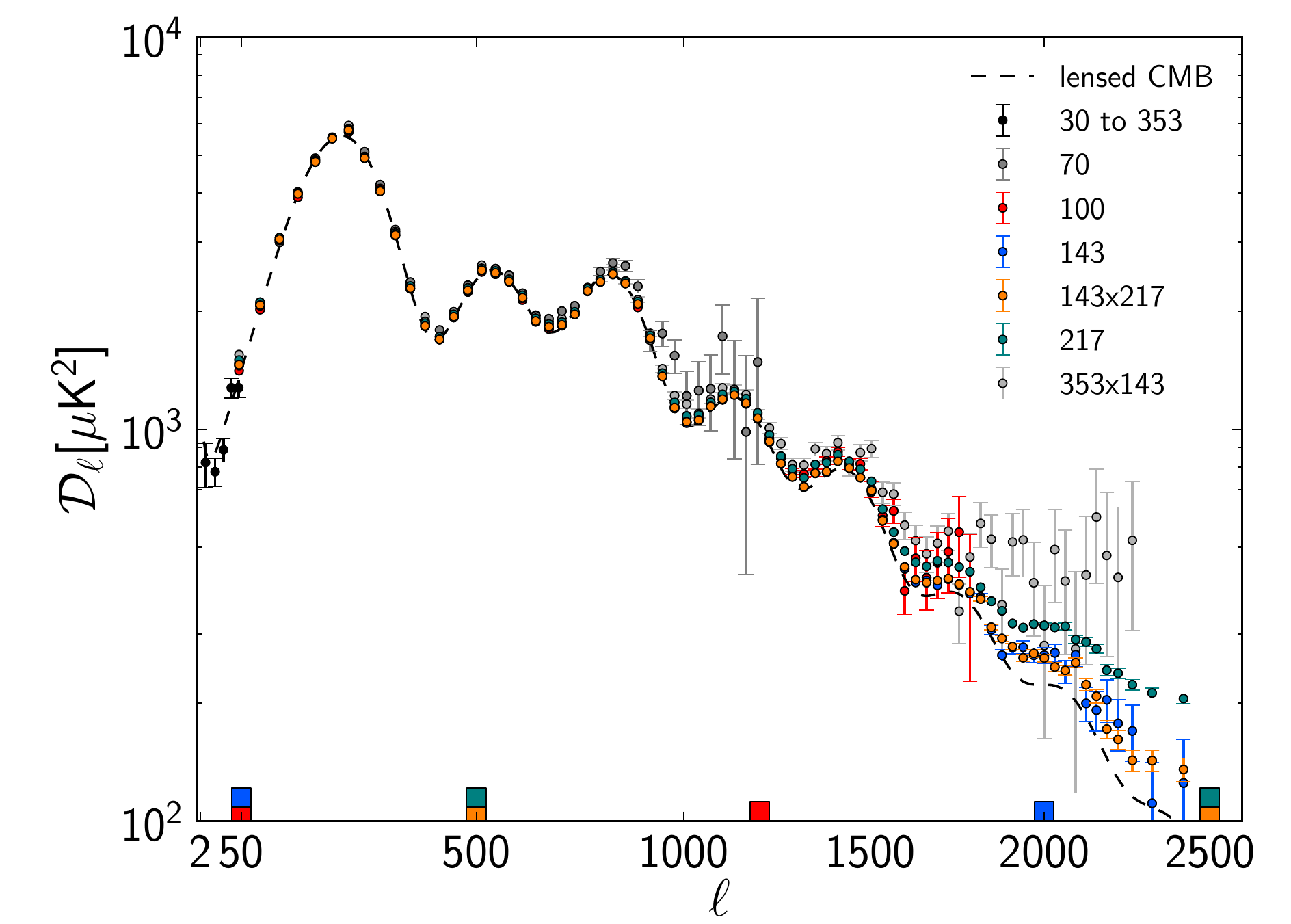}
   \caption{Power spectrum of cosmic microwave background radiation measured by Planck collaboration
\cite{Ade:2013kta}.} 
  \label{fig_planck}
  \end{center}
\end{figure}

 Unlike sounds to be discussed
at
the end of this paper, here we consider sounds  propagating in  Universe at much later time,  when
the primordial plasma gets neutralized into atoms.  The corresponding temperature
was of the order of an electron-Volt,  12 orders lower than in electroweak and 9 orders lower than
in QCD phase transitions. It is at this stage of Big Bang at which photons which we now see as   cosmic microwave
background radiation were emitted. These sounds lead to famous deviations of the background radiation
temperature, of magnitude $10^{-5}$, from the mean $T$ of the Universe.
 The data by Planck collaboration on their angular harmonics power spectrum (distributed over the sky $\theta,
 \phi$ angles)
 of these 
   perturbations are   shown in Fig. \ref{fig_planck}.
   
   They show a dissipation toward higher harmonics, modulated by a
    number of the so called ``acoustic peaks".
 Their explanation is as follows:   since all harmonics start at the same time by Big Bang -- hydro velocities at time zero are assumed zero for all harmonics --
   and get frozen at the same time as well, they have exactly the same propagation time. 
  Their oscillation phases are however all different because 
 different harmonics have different oscillation frequencies.
Those with  larger $n$
   rotate more rapidly -- the frequency is $\sim n$. Binary correlator is proportional to $cos^2(\phi^n_{freezout})$
   and harmonics with the optimal phases close to $\pi/2$ or $3\pi/2$ etc  show maxima, maxima in between.
  
\begin{figure}[t!]
  \begin{center}
  \includegraphics[width=8cm]{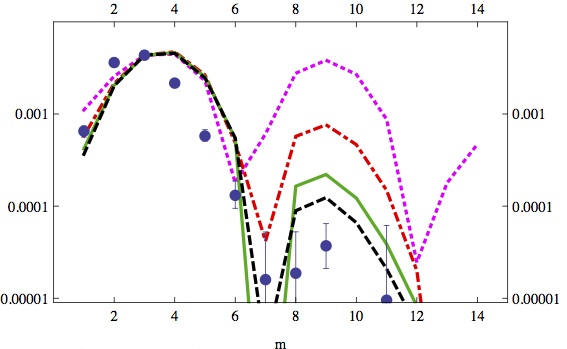}
  \includegraphics[width=9cm]{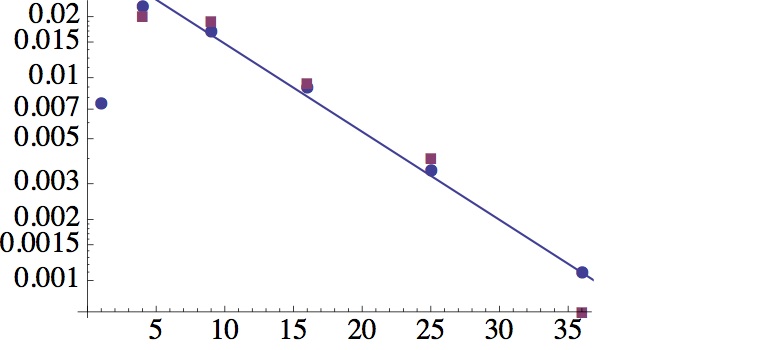} 
   \caption{(a) The lines are hydro calculations of the correlation function harmonics, $v_m^2$, based on 
   a Green function from a point source \cite{Staig:2011wj} for four values of viscosety  $4\pi\eta/s$ =0,1,1.68,2 (top to bottom at the right). The closed circles are the Atlas data
   for the ultra-central bin.
(b) $v_n\{2\}$ plotted vs $n^2$.
Blue closed circles are calculation of  via viscous even-by-event hydrodynamics \cite{Rose:2014fba},
``IP Glasma+Music",  with $\eta/s=0.14$. The straight line, shown  to guide the eye,
demonstrate that ``acoustic systematics" does in fact describe the results of this heavy calculation
quite accurately. The
CMS data for the 0-1\% centrality bin, shown by the red squares, in fact display  larger deviations,
perhaps an oscillatory ones.
   }
  \label{fig_vn_comp}
  \end{center}
\end{figure}

 At this point the curious reader would probably ask,
 if the power spectrum of harmonics do show similar   oscillations for the Little Bang as well?
 In fact in our hydro calculation we do see them in hydro calculations described above:
 with the peak around $m=3$ and the next 
 at $m=9$, with the minimum 
 predicted to be around $m=7$, see Fig.\ref{fig_vn_comp}(a).
 More recent sophisticated event-by-event   
 hydro calculation by Rose et al \cite{Rose:2014fba} does not reproduce oscillations around 
 the smooth sound damping trend, see Fig.\ref{fig_vn_comp}(b).
 One may think that averaging over many bumps in multiple configurations may indeed average out the
 freeze out phase factor. 
 Yet the ultra-central  data 
 one can still see clear deviation from the damping curve $\sim exp(-n^2*const)
$. In particular, the third harmonics is more robust than the second
$v_3>v_2$, while $v_6$ is lower than the curve. 
The point at $m=9$ is a one-sigma effect, not a statistically significant observation.

Let me conclude this
discussion with a statement, that unlike in the Big Bang, for the Little one we only have 
 certain hints for an oscillatory deviations from the ``acoustic systematics".  At this time one cannot claim that such oscillation do  exist, and even if  so, that they  agree or not with 
 the  theory.


\subsection{The smallest drops of QGP have sounds as well}
In the chapters above we have described 
 successes of hydrodynamics for description of the flow harmonics, resulting
 from sound waves generated by the initial state perturbations. We also emphasised
 the debate about the initial out-of-equilibrium stage of the collisions, and a significant 
 gap which still exists between approaches based on weak and strong couplings,
 in respect to equilibration time and matter viscosity. 
  Needless to say, the key to all those issues should be found in experimentations
  with systems smaller than central AA collisions. They should eventually should {\em the limits of hydrodynamics}
  and reveal what exactly happen in this hotly disputed ``the first 1 fm/c" of the collisions.
  
Let us start this discussion with another look at the flow harmonics. What spatial scale corresponds to
the highest $n$ of the $v_n$ observed, and does that shed light on the equilibration issue?
Here one should split discussion of sounds moving so to say in $\phi$ direction, along the fireball $surface$,
and those along the $radius$. 

A successful description of the $n$-th
 harmonics along the fireball $surface$ implies that hydro still works at a scale $2\pi R/n$: taken the nuclear radius $R\sim 6\, fm$ and the largest harmonic studied in hydro $n=6$ one concludes that
this scale is  still few fm. So, it is still large enough, and it is impossible
to tell the difference between the initial states of the Glauber model (operating with nucleons) from 
those generated by parton or glasma-based models  (operating on quark-gluon level) .
And indeed, 
 as we argued in detail above, we don't see harmonics with larger $n$ simply because 
 of current statistical limitations of the data sample. Higher harmonics suffer stronger
 viscous damping, during the long time to  freezeout. In short, non-observation of $v_n,n>6$
  {\em does not} reveal the limits of hydrodynamics.
 
%


Obviously,  one can observe  smaller and smaller systems,
e.g. $CuCu$ and lighter nuclei, and see what happens to flow harmonics. 
Note that in such case the time to freeze out is shorter, and $\epsilon_n$ larger, 
so one may hope to understand the sound damping phenomena more systematically.
 Monitoring of the collective phenomena in them would be extremely
valuable for answering those questions. 
However, it is not how the actual development went. 
Unexpectedly harmonic flows were found in very small systems --  $pp$ and $pA$ collisions, with certain high multiplicity trigger.
  
   \begin{figure}[t!]
  \begin{center}
  \includegraphics[width=8cm]{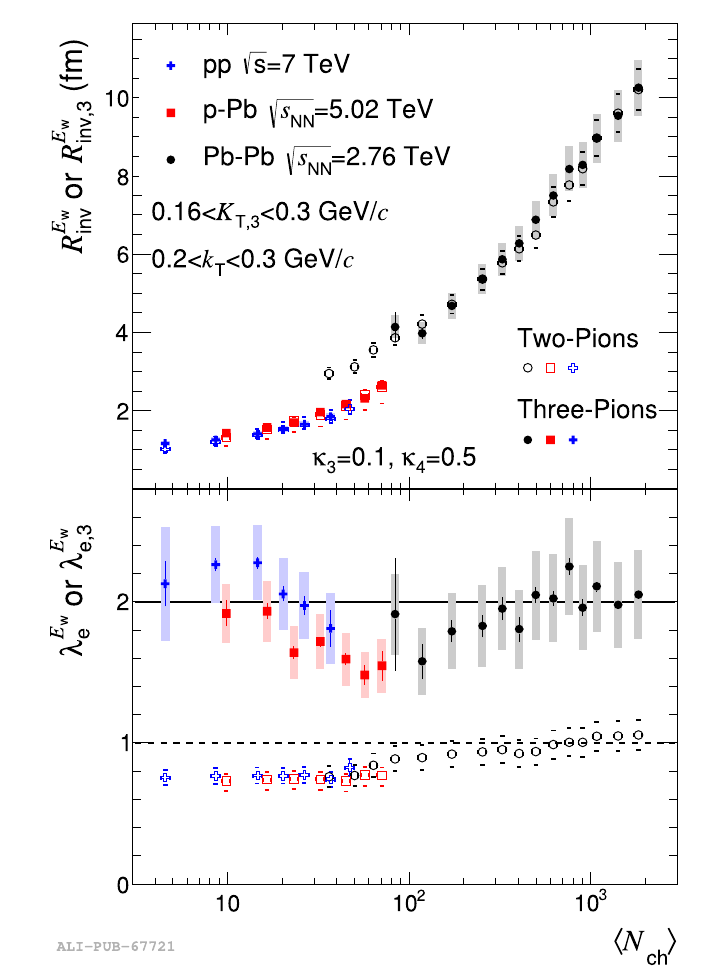}
   \caption{(From \cite{ALICE_qm20141}) Alice data on the femtoscopy radii (upper part) and ``coherence parameter" (lower part)
   as a function of multiplicity, for $pp,pPb,PbPb$ collisions. }
  \label{fig_A_HBT}
  \end{center}
\end{figure}

Before we go into details, let us try to see how large those systems really are.
At freezeout the size can be directly measured, using femtoscopy  method.
(Brief history: so called Hanbury-Brown-Twiss (HBT) radii. This interferometry method  came from radio astronomy.
The influence of Bose symmetrization of the wave function of the observed mesons in particle physics 
was first emphasized by Goldhaber et al \cite{Goldhaber:1960sf} and applied to proton-antiproton annihilation.
Its use for the determination of the size/duration of the particle production processes had been proposed 
by Kopylov and Podgoretsky \cite{Kopylov:1973qq} and myself \cite{Shuryak:1974am}. 
With the advent of heavy ion
collisions this ``femtoscopy'' technique had grew into a large industry. Early applications for RHIC 
heavy ion collisions were in certain tension with the hydrodynamical models, although this issue was later
resolved \cite{Pratt:2008qv}.)

The corresponding data are shown in Fig.\ref{fig_A_HBT}, which combines the
traditional 2-pion and more novel 3-pion correlation functions of identical pions. 
An overall growth of the freezeout size with multiplicity, roughly as $<N_{ch}>^{1/3}$,
is expected already from the simplest picture, in which the freezeout density is some universal constant.
For AA collisions this simple idea roughly works: 3 orders of magnitude of the growth in multiplicity correspond
to one order of magnitude growth of the size. 
 
 Yet the $pp, pA$ data apparently fall on a different line, with significantly  smaller radii, even 
 if compared to the peripheral AA collisions at the same multiplicity. Why do those systems 
 get frozen at higher density, than those produced in AA? 
  To understand why can it be the case
 one should recall the {\em freezeout condition}:
 ``the collision rate becomes comparable to the expansion rate"
 \be <n\sigma v> =\tau_{coll}^{-1}(n)\sim  \tau_{expansion}^{-1}= { dn(\tau)/d\tau \over n  (\tau)} \ee
Higher density means larger l.h.s., and thus we need a larger r.h.s.. So, 
we see that new ``very small Bangs" are in fact
more ``explosive", with larger expansion rate. 
We will not go into relevant data and theory, but just state that 
indeed this conclusion is supported by stronger radial flow in 
 $pp,pA$ high-multiplcity systems,  supporting directly what we just learned from
 the HBT radii.  

But $how$ those systems become ``more explosive" in the first place? Where is the room for that,
people usually ask, given that even the $final$ sizes of these objects are small?
Well, the only space left is at the beginning:  those systems must 
be born very small indeed, and start accelerating stronger, to generate
strong collective flows observed.
How it may happen is a puzzle which is now hotly debated in the field.

  \begin{figure}[h!]
\begin{center}
\includegraphics [width=6.cm]{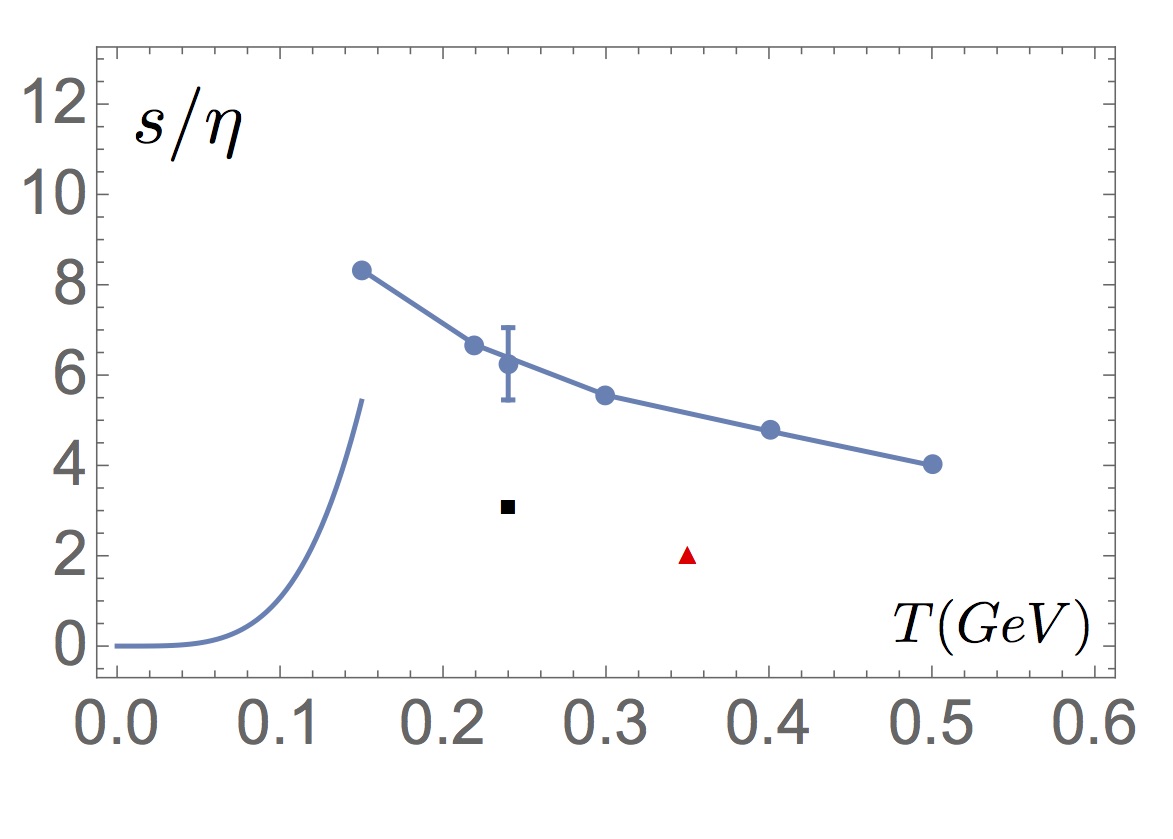}
\includegraphics[width=6.cm]{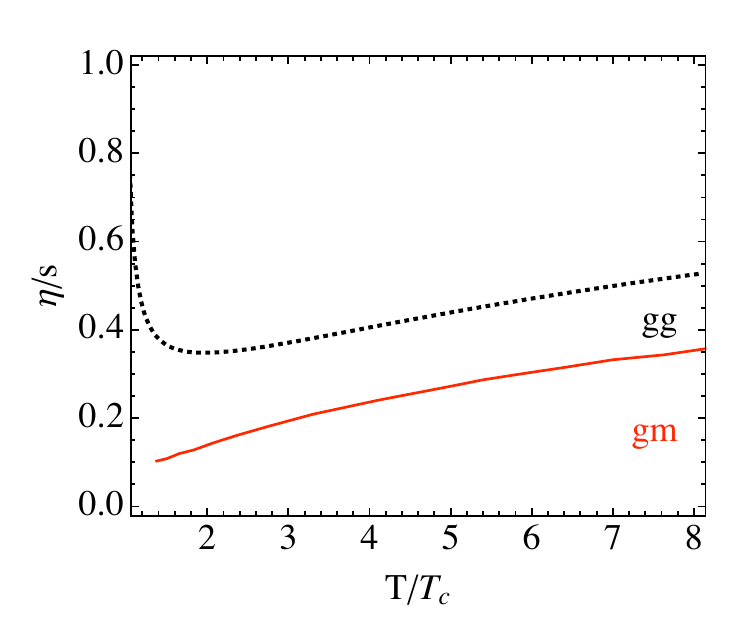}
\caption{Left plot: The entropy density to shear viscosity ratio $s/\eta$ versus
the  temperature $T\, (GeV)$. The upper range of the plot, 
$s/\eta=4\pi$ corresponds to
the value in infinitely strongly coupled $\cal{N}$=4 plasma \cite{Policastro:2001yc}. The curve without points
on the left  corresponds to hadronic/pion rescattering according to chiral perturbation theory \cite{Prakash:1993bt}.
The single (red) triangle 
corresponds to a molecular dynamics study of classical strongly coupled colored plasma
\cite{Gelman:2006xw},  the single (black) square corresponds to numerical evaluation  on the lattice \cite{Nakamura:2004sy}.
The single point with the error bar corresponds to the phenomenological value extracted from the data, see text.
The series of points connected by a line on the right side correspond to gluon-monopole scattering \cite{Ratti:2008jz}.
Right plot:  The inverse 
 ratio $\eta/s$ as a function of the temperature normalized to its
  critical value $T/T_c$. The solid line marked $gm$ corresponds to the gluon-monopole scattering \cite{Ratti:2008jz}, 
  same as in the upper plot, the dashed line shows
the perturbative gluon-gluon scattering: this line is shown for comparison.}
 \label{fig_SoverEta}
\end{center}
\end{figure}

\subsection{Why is the QGP  such an unusual fluid?}
  Multiple experiments, with heavy ions and ``smaller systems" just described above,
 allowed us to
 extract the values of kinetic coefficients, such as  shear viscosity $\eta$. 
 In a kinetic theory it is proportional to mean free path, which is inversely proportional to density of constituents and their transport cross section. The ratio of the  entropy density to it
   \be  {s\over \eta }\sim {n \sigma_{transport} \over  T \bar{v} } \ee
is basically  the ratio of interparticle separation to the mean free path.
It should be small in weak coupling (small cross section), but is in fact much larger than one,
see Fig. \ref{fig_SoverEta}.  

The density of ``electric" (quark and gluon) 
quasiparticles rapidly decrease as $T\rightarrow T_c$
since
they are eliminated by the phenomenon of electric confinement. One might then expect
the $s/\eta$ ratio to decrease as well, but in fact 
 (see Fig.\ref{fig_SoverEta})
  ${s/ \eta }$ has instead  a $peak$ there.
  This peak correlates with  similar peaks claimed for two more kinetic parameters,
 the  heavy quark diffusion constant and
  the jet quenching
  parameter $\hat{q}$.

 As $T$ decreases, toward the end of the QGP phase at $T_c$ , the effective coupling grows,
and one need to use some non-perturbative methods rather than Feynman diagrams.
Opinions differ on how one should describe matter in this domain.
Different schools of thought can be classified as (i) perturbative, (ii) semiclassical; (iii) dual magnetic;
and (iv) dual holographic ones.

 What can be called ``the  semiclassical direction" focuses on 
evaluation of the path integral over the fields using generalization of the saddle point method. The
 extrema of its integrand are identified and their contributions evaluated. 
 It is so far most developed in quantum mechanical models, for which 2 and even 3-loop
 corrections have been calculated. In the case of gauge theories extrema are ``instantons",  
 complementing perturbative series by terms $\sim exp(-const/g^2)$ times the so called ``instanton series" in $g^2$.
This result in the so called {\em trans-series},  which are not only more accurate
than perturbative ones, but they are suppose to be free from
ambiguities and unphysical imaginary parts, which perturbative and instanton series have
separately.

For the finite-temperature applications, plugging logarithmic running of the coupling into such exponential
terms  one finds some
  {\em power} dependences of the type
\be e^{-S}\sim exp\left(-{const\over g^2(T)}\right) \sim ({\Lambda \over T})^{power} \ee 
So, these effects are
not important at high $T$ but explode -- as inverse powers of $T$ -- near $T_c$. 
%

In 1980-1990's it has been shown how instanton-induced interaction between light quarks
 {\em break the chiral symmetries}, the $U_A(1)$ explicitly and $SU(N_f)$ spontaneously. The latter is understood
via collectivization of fermionic zero modes, for a review see \cite{Schafer:1996wv}. Account for non-zero average
Polyakov line, or non-zero vacuum expectation value
of the zeroth component of the gauge potential $<A_0>$  require re-defined solitons, in which this gauge field component does not vanish at large distances. 
Account for this changed instantons into  
  a set of $N_c$ instanton constituents,  the so called Lee-Li-Kraan-van Baal (LLKvB) {\em instanton-dyons}, or instanton-monopoles \cite{Lee:1998bb,Kraan:1998sn}.  It has been recently shown that those, if dense enough, can  naturally generate $both$ confinement
and chiral symmetry breaking, see \cite{Liu:2015jsa,Larsen:2015tso}, for recent review see \cite{Shuryak:2016vow}.
These works are however too recent to have impact on heavy ion physics, and we will not discuss them here.

(iii) A "dual magnetic" school consists of two distinct approaches. A ``puristic"
point of view assumes that at the momentum scale of interest the electric coupling is
 large, $\alpha_s \gg 1$, and therefore there is no hope to progress with the usual ``electric" formulation of the gauge theory,
 and therefore one should proceed with building its
  ``magnetic" formulation,
with weak ``magnetic coupling" $\alpha_m=1/\alpha_s\ll 1$.  
Working example of effective magnetic theory of such kind were demonstrated 
 for supersymmetric theories, see e.g. \cite{Seiberg:1994rs}.  For applications of the dual magnetic model to
 QCD flux tubes see \cite{Baker:1997bg}.
 
 A more pragmatic  point of view -- known as ``magnetic scenario" -- starts with acknowledgement
 that both electric and magnetic couplings are close to one, $\alpha_m\sim \alpha_e\sim 1$. So,
  neither perturbative/semiclassical nor dual formulation will work quantitatively.
   Effective masses, couplings and other properties of all coexisting quasiparticles -- quarks, gluons and
  magnetic monopoles  -- can only be deduced phenomenologically, from the analysis of
 lattice simulations. 
  We will discuss this scenario below in this section.

(iv)  Finally, very popular during the last decade are ``holographic dualities",  connecting 
  strongly coupled gauge theories to a string theory in the curved space  with extra dimensions.
As shown by \cite{Maldacena:1997re}, in the limit of the large number of colors, $N_c\rightarrow \infty$, 
it is a duality to much simpler -- and
weakly coupled  -- theory,  a modification of {\em classical gravity}.  Such duality 
 relates problems we wish to study
   ``holographically" to some problems in general relativity. In particular, the
thermally equilibrated QGP at strong coupling is related to  certain black hole solutions in 5 dimensions,
in which the plasma temperature is identified with the Hawking temperature, and the QGP entropy 
with the Bekenstein entropy.  

Completing this round of comments, we now return to  (iii), the approach focused on magnetically charged quasiparticles, and provide more details on its  history, basic ideas and results.  

Already J.J.Thompson, the discoverer of the electron,  
 noticed that something unusual should happen already for static electric and magnetic charges existing together.
While both the electric field $\vec E$ (pointing from the center of the electric charge $e$) and the magnetic 
field $\vec B$ (pointing from the center of the magnetic charge $g$) are static (time independent), the Pointing vector
$S=[\vec E\times \vec B]$ indicates that the energy of the electromagnetic field  rotates
around the line connecting the charges.

A.Poincare went further, allowing one of the charges  to
move in the field of another. The Lorentz force 
\be m \ddot {\vec{r}}=- e g {[\dot{ \vec{r}} \times \vec{r} ]\over r^3} \ee
is  proportional to the product of two charges, electric $e$ and magnetic $g$.
The total angular momentum of the system has a Thompson term, also with such product
\be \vec{J}=m [\vec{r} \times\dot{ \vec{r}}] +eg {\vec{r} \over r} \ee 
Its  conservation leads to unusual consequences: unlike for the usual potential forces, in this case the
particle  motion is not restricted to the  scattering plane, normal to $\vec J$,  but to a different surface,  the {\em Poincare cone}. 

The quantum-mechanical version of this problem, involving a pair of electrically and magnetically charged particles, provides further surprises.
The angular momentum of the field mentioned above must take values proportional to $\hbar$ with integer or semi-integer coefficient: this leads to famous
{\em Dirac quantization condition} \cite{Dirac:1931kp} 
 \be   e g ={1\over 2} \hbar c  n 
 \ee  
(where we keep $\hbar$, unlike most other formulae) 
with an integer $n$ in the r.h.s. Dirac himself derived it differently, 
arguing that the unavoidable singularities of the gauge potential of the form of the Dirac strings should be pure gauge artifacts and thus invisible. He emphatically noted that this relation was the first suggested reason in theoretical literature for the electric charge
quantization. If there be just one monopole in Universe, all electric charges $must$
obey this relation, or electrodynamics gets inconsistent with quantum theory!

Many outstanding theorists -- Dirac and Tamm among them -- wrote  papers about a quantum-mechanical
version of the quantum-mechanical problem of a monopole in a field of a charge, 
yet this problem was fully solved
only decades later  \cite{Schwinger:1976fr,Boulware:1976tv}. 
It is unfortunate that this beautiful and instructive problem is not -- to our knowledge -- part
of any textbooks on quantum mechanics. The key element was substitution the usual 
angular harmonics $Y_{l,m}(\theta,\phi)$ by other functions, which for large $l,m$ 
replicates the Poincare cone (rather than the scattering plane).

The resurfaced interest to monopoles in 1970s was of course inspired by the discovery 
of 't Hooft-Polyakov 
  monopole solution \cite{'tHooft:1974qc,Polyakov:1974ek} 
  for Georgi-Glashow model, with an adjoint scalar field complementing the non-Abelian gauge field. 
Can such 
 monopoles be  quasiparticles in QGP? A confinement mechanism 
conjectured in \cite{Mandelstam:1974pi,'tHooft:1977hy}
suggested that spin-zero monopoles may undergo
a Bose-Einstein condensation, provided their density is large enough and the temperature sufficiently low.
 These ideas, known as the ``dual superconductor" model, were strongly supported by lattice studies, in which one can detect monopoles and do see how those make a
 ``magnetic current coil" stabilizing the electric flux tubes. 
 

The monopole story continued  at the level of quantum field theories (QFTs), with another fascinating turn. 
Dirac considered the electric and magnetic charges $e,g$ to be some parameters,
defined at large distances from the charges.  
But in QFTs the charges run as a function of the momentum scale,
as prescribed by the renormalization group (RG) flows. So, we came
to important realization: in order to keep the
Dirac condition valid at all scales,   $e(Q)$ and $g(Q)$ must be
running {\em  in the opposite directions}, keeping their product fixed!

In QCD-like theories, with the so called asymptotic freedom,
 the electric coupling is small in UV (large momenta $Q$ and temperature $T$) and increases toward the
 IR (small $Q$ and $T$). 
%
%
%

How the electric and magnetic RG flows work has been first demonstrated
by a great example,  the $\cal{N}$=2 supersymmetric  theory,
for which the solution was found by Seiberg and Witten in \cite{Seiberg:1994rs}. In this theory the monopoles do exist
as particles, with well-defined masses. 
When the vacuum expectation value (VEV) of the Higgs field is large, there is weak
(perturbative)  regime for electric particles, gluinoes and gluons.
In this limit monopoles are heavy and strongly interacting. However,
for  certain 
special values of VEV,  they  do indeed
become light and weakly interacting, while the electric ones -- gluons and gluinos -- are very heavy and strongly interacting.
The corresponding low energy magnetic theory
is  nothing else but the (supersymmetric version of)  QED, and 
its beta function, as expected, has the opposite sign to that of the electric theory.

Even greater examples are provided by the 4-dimensional conformal theories, such as
 $\cal{N}$=4 super-Yang-Mills.  Those theories are {\em electric-magnetic selfdual}. This means that monopoles, dressed by all
fermions bound to them, form the same supermultiplet as the original fields of the ``electric theory". Therefore,  the beta function of this theory should be equal to  itself with the
minus sign! The only solution to that requirement is that the beta function must be identically zero,  the is no running of the coupling at all, the theory is conformal.



Completing this brief pedagogical update, let us return to \cite{Liao:2006ry} paper, 
considering properties of a
classical plasma, including both electrically and magnetically charged particles.
 Let us proceed in steps of complexity of the problem, starting from 3 particles: a
 $pair$ of $\pm q$ static electric charges, plus a monopole which can move in their ``dipole field". 
  Numerical integration of the equation of motion had showed that
  the monopole's motion takes place on  a curious surface, interpolating two  Poincare cones with ends 
at the two charges: so-to-say, two charges play ping-pong with a monopole, without even moving!
Another way to explain it is by noting that an electric dipole is ``dual" to a ``magnetic bottle", with magnetic coils,
 invented  to
 keep electrically charged particles inside.
 
 The next example was a cell  with 8 alternating static
 positive and negative charges --  modeling a grain of salt.  A monopole, which is initially placed inside the cell,
 has
   formidable obstacles   to get out of it: hundreds of scattering with the corner charges happen 
   before it takes place. 
 The Lorentz force acting on magnetic charge forces it to rotate around the electric field. Closer to the charge
 the field grows and thus rotation radius decreases, and eventually two particles collide.
 
 Finally, multiple (hundreds) of electric and magnetic particles were considered  in  \cite{Liao:2006ry}, 
 moving 
 according to classical equation of motions. It was found that their paths essentially
replicate the previous example, with each particle being in a ``cage",
  made by its dual neighbors.   These findings provide some explanation  of why  electric-magnetic plasma has unusually 
 small mean free path and, as a result, an unusually perfect
  collective behavior.


At the
  quantum-mechanical level the many-body studies of such plasma are still to be done. So one has to rely on
  kinetic theory and 
  binary cross sections. Those for
  gluon-monopole scattering were calculated in
\cite{Ratti:2008jz}. It was found that gluon-monopole scattering dominates over the gluon-gluon one, as far as  transport cross sections are concerned. 
and produce values of the viscosity quite comparable with that is observed in sQGP experimentally, as was already shown in Fig.\ref{fig_SoverEta} . What is also worth noting, it does predict a maximum of this 
ratio at $T=T_c$, reflecting the behavior of the density of monopoles.

%

Returning to QCD-like theories which  do not have powerful extended supersymmetries which would
prevent any phase transitions and guarantee smooth transition from UV to IR,
one finds transition to confining and chirally broken phases. Those have certain quantum condensates
which divert the RG flow to hadronic phase at $T<T_c$. Therefore 
the duality argument must hold at least in the plasma phase, at $T>T_c$. We can follow the 
duality argument and the Dirac condition only half way, till $e^2/4\pi\hbar c\sim g^2/4\pi\hbar c\sim 1$.
This is a plasma of coexisting electric quasiparticles and magnetic monopoles.

 \begin{figure}[t]
  \begin{center}
  \includegraphics[width=8cm]{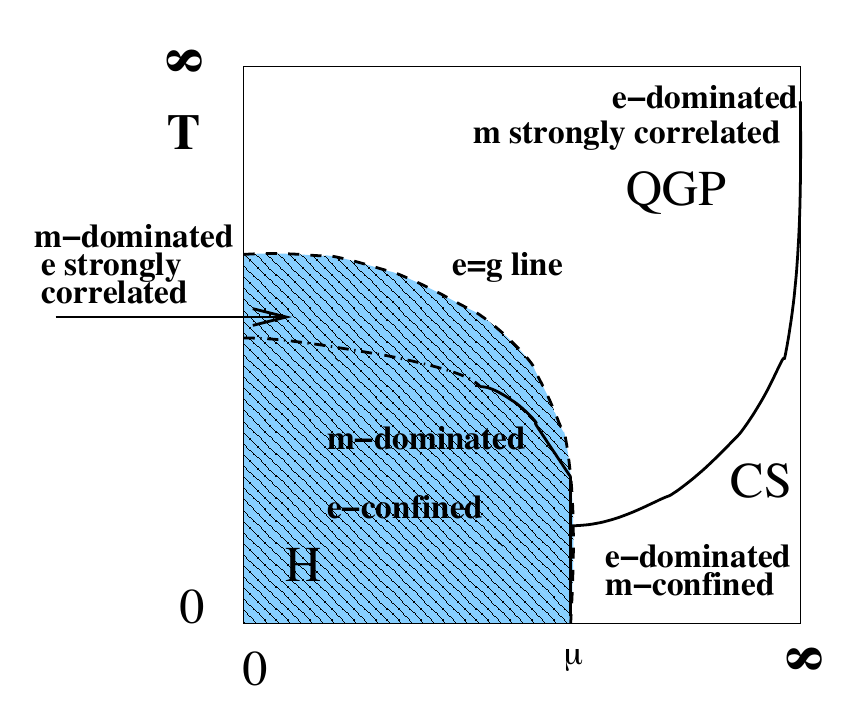}
   \caption{ A schematic phase diagram on a (``compactified") plane of temperature and baryonic chemical potential,  $T - \mu$, from \cite{Liao:2006ry}. The (blue) shaded region shows ``magnetically dominated" region $g < e$, which includes the deconfined hadronic phase as well as a small part of the QGP domain. Unshaded region includes the ``electrically dominated" part of QGP and the color superconducting (CS) region, which has e-charged diquark condensates and is therefore ``magnetically confined". The dashed line called ``e=g line" is the line of electric-magnetic equilibrium. The solid lines indicate true phase transitions, while the dash-dotted line is a deconfinement cross-over line.
}
  \label{fig_EM_duality}
  \end{center}
\end{figure}

 One can summarize the picture of the so called ``magnetic scenario" by a schematic plot shown in
Fig. \ref{fig_EM_duality}, from \cite{Liao:2006ry}.  At the top -- the high $T$ domain -- and at the right -- the high density
domain -- one finds 
weakly coupled or ``electrically dominated" regimes, or wQGP. On the contrary, near the origin
of the plot, in vacuum,   the electric fields are, 
subdominant and confined into the flux tubes. The vacuum is filled by the magnetically charged condensate,
known as ``dual superconductor". The region in between (relevant for matter produced at RHIC/LHC) is 
close to  the ``equilibrium line", marked by $e=g$ on the plot. (People for whom couplings are too abstract, can for example 
define it by an equality of the electric and magnetic screening masses.) In this region both electric and magnetic coupling
are equal and thus $\alpha_{electric}=\alpha_{magnetic}=1$: so neither the electric nor magnetic formulations of the theory are simple. 

Do we have any evidence for a presence or importance for heavy ion physics of ``magnetic" objects? Here are some arguments for that
based on lattice studies and phenomenology, more or less in historical order:

(i) In the RHIC/LHC region $T_c<T<2T_c$ the VEV of the Polyakov line $<P>$ is substantially different from 1.  It was argued by \cite{Hidaka:2008dr}  that $<P>$ must be incorporated into density of thermal quarks and gluons, and thus
suppress their  contributions.  They called such matter ``semi-QGP"
emphasizing that say only about half of QGP  degrees of freedom  should actually contribute to thermodynamics
at such $T$. 
And yet, the lattice data insist that the
  thermal  energy density normalized as $\epsilon/T^4$ remains constant nearly
  all the way to $T_c$. 
  
  (ii) ``Magnetic scenario" \cite{Liao:2006ry} proposes to
 explain this puzzle by ascribing ``another half" of such contributions to the magnetic monopoles, which
  are not subject to  $<P>$ suppression because they do not have the electric charge. 
  A number of lattice studies found magnetic monopoles and 
 had shown that they behave as physical quasiparticles in the medium. Their motion definitely shows Bose-Einstein condensation at $T<T_c$ \cite{D'Alessandro:2010xg}. Their spatial correlation
functions are very much plasma-like. Even more striking is
the observation \cite{Liao:2008jg} revealing magnetic coupling which $grows$ with $T$, being indeed an
inverse of the asymptotic freedom curve. 

The magnetic scenario also has difficulties. Unlike instanton-dyons we mentioned,
lattice monopoles so far defined are gauge dependent.  The original 'tHooft-Polyakov solution
require an adjoint scalar field, absent in QCD Lagrangian, but perhaps an effective scalar can be generated dynamically. 
In the Euclidean time finite-temperature setting this is not a problem, as $A_0$ naturally takes this role,
but  it cannot be used in real-time applications required for kinetic calculations. 

(iii) Plasmas with electric and magnetic charges show unusual transport properties: Lorenz force enhances
  collision rate and reduce viscosity \cite{Liao:2006ry}. Quantum gluon-monopole scattering leads to large transport cross section \cite{Ratti:2008jz},
  providing small viscosity in the range close to that observed at RHIC/LHC.
  
 (iv) The high density of (non-condensed) monopoles near $T_c$ leads to compression of the electric flux tubes, perhaps explaining   curious
 lattice observations of very high tension in the potential energy (not free energy)
 of the heavy-quark potentials near $T_c$ \cite{Liao:2006ry}.

(v) Last but not least, the peaking density of monopoles near $T_c$ seem to be directly relevant to
jet quenching.

Completing this introduction to monopole applications, it is impossible not to mention the remaining unresolved issues.
Theories with adjoint scalar fields -- such as e.g. celebrated $\cal{N}$=2 Seiberg-Witten theory -- naturally have
particle-like monopole solutions. Yet in QCD-like theories without scalars the exact structure of
the lattice monopole are not yet well understood.

    \section{Are cosmological phase transitions observable?}     \label{sec_cosmo} 
    Since this review is aimed at non-specialists, 
some introductory information about the cosmological phase transitions
is included in Appendix \ref{sec_app_cosmo}.

Admittedly, the question in the title of this section is too general: there are many ways in which
electroweak and QCD transitions may affect present day Universe. For example, 
electroweak transitions must be crucially important for the baryon asymmetry of the Universe.
We of course will discuss only one possible answer to it, related with gravitational waves.

\begin{figure}[htbp]
\begin{center}
\includegraphics[width=10cm]{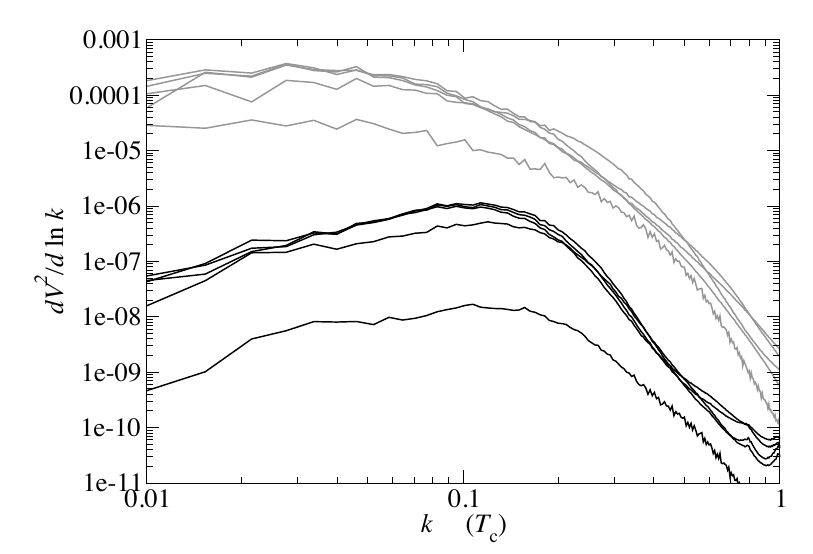}
\caption{From \cite{Hindmarsh:2013xza}. Power spectrum of the velocity squared
versus the (log of) the wave number $k$.  The grey upper
curves are for sounds, from down up as time progresses,for 
$t T_c= 600, 800, 1000, 1200, 1400$. The
black
curves
in the
 bottom are for rotational excitations.}
\label{fig_sound_spectrum_ew}
\end{center}
\end{figure}

\subsection{Sounds from the phase transitions} 
%

 We think that our Universe has been ``boiling" at its early stages (at least) three times: (i) at
the initial  equilibration, when entropy was produced, at (ii) electroweak and (iii) QCD phase transitions.
On general grounds, these should have produced certain out-of-equilibrium effects, resulting
in inhomogeneuities and thus sound. (As an example well familiar to anyone,
recall that a cattle start ``singing" as tea is ready. The critical phenomenon is production of
vapor bubbles, which then collapse and pass their energy to sounds.)

Theoretical studies of this process, both for electroweak and QCD transitions, are carried out
for at least three decades. An example of such calculation for electroweak transition is shown in Fig,\ref{fig_sound_spectrum_ew} assuming the transition is of the first order. One lesson from it is that the sounds (upper grey curves) dominate the rotations (lower black curves).
Another impressive result is that the simulation was able to cover two orders
of magnitude of the wavelengths. And yet, there are many more decades of $k$
to the left of this plot which needs to be explored, before we reach the IR cutoff of the process, the scale at which we hope to observe gravity waves.

Experiments with heavy ion collisions, which do create passing through $T_c$ and do observe
sounds (as we discussed already above). And yet, those sounds so far observed originate from
inhomogeneous initial conditions, not the near-$T_c$ critical region. How it can be done
has been proposed -- e.g. in my paper with Staig \cite{Shuryak:2013uaa} --   
but not so far carried out. 

Yet sound production is not the main issue here, the fate of subsequent {\em acoustic cascade} is.
The main proposal of our paper \cite{ourGW} is that it can go into a regime known as 
{\em inverse acoustic cascade}. If it does, the sounds created at the thermal scale
can get hugely amplified toward the IR scale. In simpler terms, it is possible that
a huge storm may develop, with a cutoff only at the scale of Universe horizon.
At the time of QCD transition, this scale is 18 orders of magnitude different from the thermal scale.

Earth atmosphere is basically 2-dimensional, its hight is three orders of magnitude smaller than Earth's diameter, and that is why the inverse cascades
create large storms. The amplification rate can be truly huge. The Universe is 3-dimensional, and in this case it can appear only in very special  circumstances.
It remains a great challenge to figure out whether it is the case, maybe for one of the transitions.

 The challenge is to understand when and how the can be developed.
 The answer, first of all, crucially depend 
on the $sign$ of small 
 corrections to sound dispersion, which we write as
 \be \omega=c_s k (1+A k^2+...) \ee 
 The sign of the correction constant $A$ is 
  not known, both for QGP   and electroweak plasma.
If $A>0$ the phonon decays $1 \rightarrow 2$ are possible. The turbulent cascade
 based on such 3-wave interactions was shown to develop in the $direct$ -- that is large $k$ or UV -- direction, which is $not$ the one
 we are interested in.

Another alternative, when the dispersive correction coefficient $A<0$ is
 negative, turns out to be much more interesting. In this case 
 the cascade switches to higher order processes,  of 
$2\leftrightarrow 2$ scattering  and/or $1\leftrightarrow 3$ processes.
The analysis of corresponding acoustic cascade
is much more involved   but it does show existence of the inverse cascade, with
a particle flow directed to IR,
with the weak turbulence index of the density momentum distribution
 \be n_k\sim k^{-s},\,\,\,\,  s_{weak}=10/3 \ee  
Furthermore, as discussed in
\cite{ourGW}, large value of the density at small $k$ leads to violation of weak turbulence applicability
condition and the regime is known as ``strong turbulence"
in which case the evaluated index is even larger, $s_{strong}= 4-6$.
This is an interesting and complex problem, since the sound-sound scattering
processes are not simple. It 
  can and should be numerically simulated, like it was done
  for scalar fields and gluonic cascades, but it was not  studied yet.


 \subsection{From the sounds to  the gravitational waves}
Before we come into more technical discussion, let us briefly note why
do we need to focus on such (still rather exotic) observable. 
Gravitational waves, as a cosmology tool, looked as a science fiction for about a century,
but not anymore, due to recent LIGO observations.

  From the onset of the QGP physics in heavy ion collisions a specially important  role has been attributed to 
the ``penetrating probes", which for heavy ion collisions mean  photons/dileptons \cite{Shuryak:1978ij}.
So it is quite logical to think also about the only   ``penetrating probe" of the Big Bang, the {\em gravity waves} (GW). 

30 years ago Witten \cite{Witten:1984rs} had discussed the cosmological QCD phase transition,
assuming it to be  of the first order: he pointed out 
 bubble production and coalescence,  producing inhomogenuities in energy distribution
and mentioned production of the gravity waves. 
Among papers followed it were estimates of how much gravity waves will be produced.

 Jumping many years to recent time, the fascinating observation was made by
 Hindmarsh \textit{et al} \cite{Hindmarsh:2013xza}. These authors calculated 
 gravity wave production, by numerically evaluating a
 correlator
 of two stress tensors
 $ < T_{\mu\nu}(x) T_{\mu\nu}(y)> $
 during the electroweak transition.  They followed phase transitions till its end,
 and obtained the sound spectra already shown above.
 During the time of the simulation, the Higgs value does settle to its eternal value
 and no changes are seen in electroweak sector any more. 
 And yet, the calculated rate of  gravity wave production
 has shown no sign of disappearing, all the way to the end of the simulation!
 
  It turned out that the dominant  source of the GW in those simulations are
 hydrodynamical sound waves. Furthermore, the GW generation rate remains constant even {\em long after} the phase transition itself is over.  So, we argued  \cite{ourGW} , there must be some acoustic cascade
 involved, since only large wavelength small-$k$ sounds can survive viscous losses for a long time.
 
In that work \cite{ourGW} we discussed the sound-based GW production further. We  argue  that generation of the cosmological GW  can be divided into four distinct
stages, each with its own physics and scales. We will list them starting from the  UV end of the spectrum
 $k\sim T$ and ending at the IR end of the spectrum $k\sim 1/t_{life}$ cutoff by the Universe lifetime at the era
: \\(i) the production of the sounds   \\(ii)
the inverse cascade" of the acoustic turbulence, moving the sound from UV to IR \\
(iii)   the final transition from sounds  to GW.

The stage (i) remains highly nontrivial, associated with the dynamical details of the
 QCD and electroweak (EW) phase transition. 
 The  stage (ii), on the other hand, is in fact amenable to perturbative studies
 of the acoustic cascade, which is governed by Boltzmann equation. It has been already rather
 well studied in literature on turbulence, in which power attractor solutions has been identified.
 Application of this theory allows to see how small-amplitude sounds can be amplified, as one goes to smaller $k$.

 The stage (iii)  can  be treated via a simple approximation allowing to calculate the  correlator
 of two stress tensors. In hydrodynamic approximation stress tensor contains $T_{\mu\nu}\approx (\epsilon+p)u_\mu u_\nu$
 where the first bracket contains the energy density and pressure of the medium, and $u_\mu$
 is 4-velocity of its motion. If one  $u_\mu$ is produced by one sound wave, and the second by another,
 one finds that the
 standard loop diagram for  the correlator splits into a square of the amplitude describing
 new elementary process:
$$ sound+sound \rightarrow graviton $$
There is no place here for technical discussions, and we  only comment on the kinematics of the process. The speed of sound $c_s\approx 1/\sqrt{3}$ is only about half speed of light,
so to get enough energy for a graviton two sounds need to cancel half of their momenta:
in a symmetric case the angle between them should be about 100$^o$ or so. 

 Finally, let us briefly touch the question whether and how the gravitational waves can be 
 detected experimentally. In appendix B we estimate the corresponding period expected
 from electroweak and QCD transitions. They are much much longer
 than those observed by LIGO (micro-seconds). 
 
 GW from the electroweak era are expected to have periods of $hours$: those will
 be searched for by future  GW observatories in space, such as eLISA.

  The GW from the QCD transition  are expected to have periods of about a $year$. 
It turns out that for that time window there exists a very nice method as well: possible
  observational tools for them are the correlations of the millisecond pulsar 
  signal coming from different direction. The basic idea is that
when GW is falling on Earth and, say, stretches distances in a certain direction, then in the orthogonal direction one expects
distances to be reduced. The binary correlation function for the pulsar time delay is an expected function of the angle $\theta$ between them on the sky. There are existing collaborations -- North American Nanohertz Observatory for Gravitational Radiation, European Pulsar Timing Array (EPTA), and Parkes Pulsar Timing Array -- which actively pursue both the search for new millisecond pulsars and collecting the timing data for some known pulsars. It is believed that about 200 known millisecond pulsars constitute only about 1 percent of the total number of them in our Galaxy.
The current bound  on the GW fraction of the energy density of the Universe
is approximately 
\begin{align}
\Omega_{GW} (f\sim 10^{-8} \mathrm{Hz}) h_{100}^2<  10^{-9}\,.
\end{align}
Rapid progress in the field, including better pulsar timing and  formation of a global collaborations of observers, is expected
to improve the sensitivity of the method , perhaps making it possible  to  detect
GW radiation, either from merging supermassive black holes (everyone is expecting to find now)
and perhaps even some stochastic background coming from the QCD
 Big Bang phase transition
we discuss.

\section{Summary}
This paper covers two fields, which are at very different stage of their development.

Heavy ion community is now dominated by large-scale experiments at two colliders,
RHIC and  LHC. We did observed the production of new form of matter, sQGP, 
followed by rapid explosion, the Little Bang. Many details of it are rather 
well studied. Not only the average behavior  is recorded and explained, but also
its event-by-event fluctuations. Small point-like perturbations lead to the ``sound circles",
observed in great details for a number of harmonics. The unusual kinetic properties
of sQGP are quantified, and explained by a number of approaches.
We discussed one of them, blaming short mean free path on peculiar magnetically charged
quasiparticles, the monopoles, copiously present in QGP near its critical temperature.

In connection to the central issues of this paper, the observability of cosmic phase transitions at Big Bang, basically
two things remain to be done. One, in heavy ion collisions, is
to detect sounds originating from the QCD phase transition era (rather than from the initial 
state perturbation, as it has been described above). The other is to figure out 
details of the sound dispersion curve, since we would like to know whether
sound waves can or cannot decay. 

In the case of electroweak plasma at its critical temperature there are obviously no laboratory experiments. But in this case
 the coupling is weak, and thus sall questions can perhaps be studied theoretically.

The cosmology community related to QCD and electroweak phase transitions
is just making its first steps. At this stage, one 
needs to develop even qualitative understanding of the relevant
acoustic turbulence regime. Depending on the particular scenario realized, the expected
magnitude of gravity waves varies by many orders of magnitude.
Perhaps some of scenarios are already excluded by the pulsar correlation data.
As for the electroweak transition, the decisive experiment are space gravity wave detectors like eLISA. Their sensitivity is so far tuned to black hole merger events, not so far to a
random  background of gravity wave we discuss. A lot of work is ahead.

\appendix

\section{Heavy ion terminology}
{\bf ``Ion"} in physics refers to atoms with some of its electrons missing.
While at various stages of the acceleration process the degree of ionization varies,
all of it is unimportant for the collisions, which always are done with nuclei fully stripped. 

By {\bf ``heavy ions"} we mean gold $Au^{197}$ (the only stable isotope in natural gold, and a
favorite of BNL) or lead  $Pb^{208}$ (the double magic nucleus used at CERN). Some experiments
with uranium $U$ has been also done, but not because of its size but rather due to its strong deformation.

{\bf Collision centrality} in physics is defined usually via an {\em impact parameter} $b$,
the minimal distance between the centers of two objects. It is a classical concept, and in quantum mechanics channels with $integer$ angular momentum $l=L/\hbar$ (in units of Plank constant)  are used. However, collisions at
very high energy have high   angular momentum and uncertainty in $b$ is small. Standard way
of thinking about centrality is to divide any observed distribution -- e.g. over the multiplicity $P_n$ --
into the so called {\em centrality classes}, histogram bins with a fixed fraction of events rather than
width. For example, many plots in the review say something like ``centrality 20-30\%":
This means that total sum $\sum_n P_n$ is taken to be 100\%, the events are split into say 10
bins, numerated 0-10,10-20,20-30 etc \%, and only events from a particular one 
 are used on the plot under consideration.
The most central bins have the largest multiplicity and are always recorded, the more
peripheral ones (say 80-100\%) often are not used or even recorded. 
While the observables -- like mean multiplicity -- decreases with centrality $b$ monotonically, 
it is not true for individual events. Multiple possible definitions
of the centrality classes may sound complicated, but it is not, and 
simple models like Glauber nucleon scattering give quite good description of all these distributions,
so in practice any centrality measure can safely be used.

{\bf The number of participant nucleons} $N_p$ plus the number of ``spectators" is the
total number of nucleons $2A$. The number of spectators (usually only the neutrons) 
propagating along the beam direction are typically recorded by special small-angle
calorimeters in both directions. Two-dimensional distributions over signals of both
such calorimeters are cut into bins of special shapes, also  
in a way that each bin keeps fixed percentage of the total.
Small corrections for nucleons suffering only small angle elastic and diffractive scatterings -- not counted as ``participants" are also made. 

{\bf Overlap region} is a region in the transverse space in which the participant nucleons are
located at the moment of the collision. Note that dues to relativistic contraction,
high energy nuclei can be viewed as purely 2-d object, with the longitudinal size reduced
by the Lorentz factor by 2-3 orders of magnitude at RHIC/LHC,
 practically to zero: therefore the collision moment is well defined and is the same for all nucleons.
 Crudely one may think of the overlap region classically, as the almond-like intersection of two
 circles, the edges of colliding nuclei. Note that its shape changes from a circle for central collisions
 to highly deformed one for peripheral collisions, at the impact parameter $b \approx 2R$.

{\bf Flow harmonics} are Fourier coefficient of the expansion in azimuthal angle $\phi$
\be {dN \over dy dp_\perp^2 d\phi}= {dN \over dy dp_\perp^2 }\left[1+ 2\sum_m v_m(p_\perp) cos(m \phi)\right] \ee 
The most important harmonics are the so called elliptic (m=2) and triangular (m=3) flows,
although there are meaningful data for $m=4,5,6$ harmonics as well.
Note that their measurements require knowing the direction of the impact parameter vector $\vec{b}$
on event-by event basis, since 
the azimuthal angle $\phi$
is counted from the $\vec b$ direction. The direction of  $\vec{b}$ and the beam define the so called {\em collision plane}.
The direction of $\vec{b}$ in transverse plane is traditionally denoted by $x$, the orthogonal direction by $y$
and the beam direction by $z$.

In practice this either comes from separate ``near beam" calorimeters, recording ``spectator" nucleons,
or from correlation with other particles. The flow harmonics are often introduced as a response on the system to the
asymmetry parameters $\epsilon_m$ describing Fourier components of matter distribution in $\phi$.
Note that $v_m$ relates to momentum distribution and $\epsilon_m$ to that in space: connection between the two
is non-trivial. 

{\bf Collectivity of flow. }
Flow harmonics were originally derived from 2-particle correlations in relative angle,
 to which they enter as mean square 
\be v_n^2\{2\} = <e^{in(\phi_1-\phi_2)}> = <|v_n|^2> \ee
Alternatively, it can be derived from multi-hadron correlation functions: for example those for 4 and 6 particles mostly used are 
\be v_n^4\{4\} = 2<|v_n|^2>^2 - <|v_n|^4> \ee
\be v_n^6\{6\} ={1\over 4}(<|v_n|^6> - 9<|v_n|^2><|v_n|^4> + 12<|v_n|^2>^3)
\ee
By ``collectivity" one mean the fact that all of such measurements produce nearly the same values
of the harmonic $$v_n\approx (v_n\{2\})^{1/2}\approx (v_n\{4\})^{1/4}\approx  (v_n\{6\})^{1/6}$$ In contrast to that, the ``non-flow" effects -- e.g. production of hadronic resonances like $\rho \rightarrow \pi \pi$ etc
-- basically  affect mostly the binary correlator $v_n\{2\}$ but not the others.
 
{\bf Soft and hard secondaries} mentioned in the main text indicate their dynamical origin.
``Soft" come from thermal heat bath, modified by collective flows, while the ``hard" ones
from partonic reactions and jet decay. The boundary is not well established
and depend on the reaction: ``soft" are with $p_\perp<4\, GeV$
while ``hard" are perhaps with  $p_\perp>10\, GeV$.

{\bf Rapidity} $y$ is defined mostly for longitudinal motion, via the longitudinal velocity being $v_z=tanh(y)$.
 There is also the so called space-time rapidity $\eta=(1/2) log[(t+z)/(t-z)]$
(which should not be mixed with viscosity, also designated by $\eta$) used in hydrodynamics.
Both transform additively under the longitudinal Lorentz boost. 

Sometimes one also uses transverse rapidity,
$v_\perp=tanh(y_\perp)$. Pseudorapidity variable is an approximate substitute for rapidity $y$, used 
when particle identification is not available.

{\bf Chemical and kinetic freezeouts} refer to stages of the explosion at which the rates of the $inelastic$ and $elastic$ 
collisions become smaller than the rate of expansion. The chemical freezeout
is called so because at this stage particle composition, somewhat resembling a chemical composition
of matter, is finalized. The kinetic or final freezeout is where the last rescattering happen: it is
similar to photosphere of the Sun or to CMB photon freezeout in cosmology.
The time-like surfaces of the chemical and  kinetic freezeouts are usually approximated by
isotherms with certain temperatures. The final particle spectrum is usually defined as
the so called Cooper-Frye integral of thermal distribution over the kinetic freezeout surface. 
\\

{\bf Femtoscopy or HBT interferometry} method  came from radio astronomy: HBT
is abbreviation for Hanbury-Brown and Twiss who developed it there.
The influence of Bose symmetrization of the wave function of the observed mesons in particle physics 
was first emphasized in \cite{Goldhaber:1960sf} and applied to proton-antiproton annihilation.
Its use for the determination of the size/duration of the particle production processes had been proposed back in 1970's
 \cite{Kopylov:1973qq,Shuryak:1974am}. 
With the advent of heavy ion
collisions this ``femtoscopy'' technique had grew into a large industry. Early applications for RHIC 
heavy ion collisions were in certain tension with the hydrodynamical models, although this issue was later
resolved, see e.g. \cite{Pratt:2008qv}.\\

\section{Cosmological phase transitions} \label{sec_app_cosmo}

 ln this section we remind for non-experts the magnitude of certain observables related to the QCD and electroweak transition. Step one is to evaluate
  redshifts of the transitions, which can be done by comparing the transition temperatures $T_c=170\, MeV$ and $T_{QCD}\sim 100\, \GeV$ with the temperature of the cosmic microwave background $T_{CMB}=2.73\, \mathrm{K}$. This leads to
  \begin{align}
  z_{QCD}= 7.6 \times10^{11},\quad z_{EW} \sim 4 \times10^{14}\,.
  \end{align}
   At the radiation-dominated era -- to which both QCD and electroweak ones belong -- the solution to Friedmann equation leads to well known relation between the time and the temperature.( Note that we use not gravitational but particle physics units, in which c=1 but the Newton constant
   $G_N=1/M_p^2$.)
   \begin{align}
   t=\left({ 90 \over 32\pi^3 N_{DOF}(t)}  \right)^{1/2} {M_P \over T^2}     \label{eqn_t}
   \end{align}
  where $M_P$ is the Planck mass and $N_{DOF}(t)$ is the effective number of bosonic degrees of freedom (see details in, e.g., PDG, Big Bang cosmology).

   Plugging in the corresponding $T$ one finds the
   the time of the QCD phase transition to be $t_{QCD}=4\times 10^{-5}\,s$, and electroweak $t_{EW}\sim 10^{-11}\,s$.
   Multiplying those times by the respective redshift factors, one finds that the $t_{QCD}$ scale today corresponds to
   about $3\times 10^7\,s=1 $  year, and the electroweak to $5\times 10^4\,s = 15$ hours.

\acknowledgments{
This work was supported in part by the U.S. Department of Energy under Contract No. DE-FG-88ER40388.
}

%





\begin{thebibliography}{-------}
\providecommand{\natexlab}[1]{#1}

\bibitem[Shuryak(2017)]{myRMP}
Shuryak, E.
\newblock {Strongly coupled quark-gluon plasma inn heavy ion collisions}.
\newblock {\em Rev. Mod. Phys.} {\bf 2017}, {\em 89},~035001.

\bibitem[Teaney \em{et~al.}(2001{\natexlab{a}})Teaney, Lauret, and
  Shuryak]{Teaney:2000cw}
Teaney, D.; Lauret, J.; Shuryak, E.V.
\newblock {Flow at the SPS and RHIC as a quark gluon plasma signature}.
\newblock {\em Phys. Rev. Lett.} {\bf 2001}, {\em 86},~4783--4786,
  \href{http://xxx.lanl.gov/abs/nucl-th/0011058}{{\normalfont
  [arXiv:nucl-th/nucl-th/0011058]}}.

\bibitem[Teaney \em{et~al.}(2001{\natexlab{b}})Teaney, Lauret, and
  Shuryak]{Teaney:2001av}
Teaney, D.; Lauret, J.; Shuryak, E.V.
\newblock {A Hydrodynamic description of heavy ion collisions at the SPS and
  RHIC} {\bf 2001}.
\newblock  \href{http://xxx.lanl.gov/abs/nucl-th/0110037}{{\normalfont
  [arXiv:nucl-th/nucl-th/0110037]}}.

\bibitem[Hirano \em{et~al.}(2006)Hirano, Heinz, Kharzeev, Lacey, and
  Nara]{Hirano:2005xf}
Hirano, T.; Heinz, U.W.; Kharzeev, D.; Lacey, R.; Nara, Y.
\newblock {Hadronic dissipative effects on elliptic flow in ultrarelativistic
  heavy-ion collisions}.
\newblock {\em Phys. Lett.} {\bf 2006}, {\em B636},~299--304,
  \href{http://xxx.lanl.gov/abs/nucl-th/0511046}{{\normalfont
  [arXiv:nucl-th/nucl-th/0511046]}}.

\bibitem[Collins and Perry(1975)]{Collins:1974ky}
Collins, J.C.; Perry, M.J.
\newblock {Superdense Matter: Neutrons Or Asymptotically Free Quarks?}
\newblock {\em Phys. Rev. Lett.} {\bf 1975}, {\em 34},~1353.

\bibitem[Shuryak(1978)]{Shuryak:1977ut}
Shuryak, E.V.
\newblock {Theory of Hadronic Plasma}.
\newblock {\em Sov. Phys. JETP} {\bf 1978}, {\em 47},~212--219.
\newblock [Zh. Eksp. Teor. Fiz.74,408(1978)].

\bibitem[Kapusta(1979)]{Kapusta:1979fh}
Kapusta, J.I.
\newblock {Quantum Chromodynamics at High Temperature}.
\newblock {\em Nucl. Phys.} {\bf 1979}, {\em B148},~461--498.

\bibitem[Bak \em{et~al.}(2007)Bak, Karch, and Yaffe]{Bak:2007fk}
Bak, D.; Karch, A.; Yaffe, L.G.
\newblock {Debye screening in strongly coupled N=4 supersymmetric Yang-Mills
  plasma}.
\newblock {\em JHEP} {\bf 2007}, {\em 08},~049,
  \href{http://xxx.lanl.gov/abs/0705.0994}{{\normalfont
  [arXiv:hep-th/0705.0994]}}.

\bibitem[Maezawa \em{et~al.}(2010)Maezawa, Aoki, Ejiri, Hatsuda, Ishii, Kanaya,
  Ukita, and Umeda]{Maezawa:2010vj}
Maezawa, Y.; Aoki, S.; Ejiri, S.; Hatsuda, T.; Ishii, N.; Kanaya, K.; Ukita,
  N.; Umeda, T.
\newblock {Electric and Magnetic Screening Masses at Finite Temperature from
  Generalized Polyakov-Line Correlations in Two-flavor Lattice QCD}.
\newblock {\em Phys. Rev.} {\bf 2010}, {\em D81},~091501,
  \href{http://xxx.lanl.gov/abs/1003.1361}{{\normalfont
  [arXiv:hep-lat/1003.1361]}}.

\bibitem[Borsanyi \em{et~al.}(2015)Borsanyi, Fodor, Katz, Pasztor, Szabo, and
  Torok]{Borsanyi:2015yka}
Borsanyi, S.; Fodor, Z.; Katz, S.D.; Pasztor, A.; Szabo, K.K.; Torok, C.
\newblock {Static $ \overline{\mathrm{Q}}\mathrm{Q} $ pair free energy and
  screening masses from correlators of Polyakov loops: continuum extrapolated
  lattice results at the QCD physical point}.
\newblock {\em JHEP} {\bf 2015}, {\em 04},~138,
  \href{http://xxx.lanl.gov/abs/1501.02173}{{\normalfont
  [arXiv:hep-lat/1501.02173]}}.

\bibitem[Hart \em{et~al.}(2000)Hart, Laine, and Philipsen]{Hart:2000ha}
Hart, A.; Laine, M.; Philipsen, O.
\newblock {Static correlation lengths in QCD at high temperatures and finite
  densities}.
\newblock {\em Nucl. Phys.} {\bf 2000}, {\em B586},~443--474,
  \href{http://xxx.lanl.gov/abs/hep-ph/0004060}{{\normalfont
  [arXiv:hep-ph/hep-ph/0004060]}}.

\bibitem[Ding \em{et~al.}(2015)Ding, Karsch, and Mukherjee]{Ding:2015ona}
Ding, H.T.; Karsch, F.; Mukherjee, S.
\newblock {Thermodynamics of strong-interaction matter from Lattice QCD}.
\newblock {\em Int. J. Mod. Phys.} {\bf 2015}, {\em E24},~1530007,
  \href{http://xxx.lanl.gov/abs/1504.05274}{{\normalfont
  [arXiv:hep-lat/1504.05274]}}.

\bibitem[Heinz and Snellings(2013)]{Heinz:2013th}
Heinz, U.; Snellings, R.
\newblock {Collective flow and viscosity in relativistic heavy-ion collisions}.
\newblock {\em Ann. Rev. Nucl. Part. Sci.} {\bf 2013}, {\em 63},~123--151,
  \href{http://xxx.lanl.gov/abs/1301.2826}{{\normalfont
  [arXiv:nucl-th/1301.2826]}}.

\bibitem[Staig and Shuryak(2011)]{Staig:2010pn}
Staig, P.; Shuryak, E.
\newblock {The Fate of the Initial State Fluctuations in Heavy Ion Collisions.
  II The Fluctuations and Sounds}.
\newblock {\em Phys. Rev.} {\bf 2011}, {\em C84},~034908,
  \href{http://xxx.lanl.gov/abs/1008.3139}{{\normalfont
  [arXiv:nucl-th/1008.3139]}}.

\bibitem[Lacey \em{et~al.}(2013)Lacey, Gu, Gong, Reynolds, Ajitanand,
  Alexander, Mwai, and Taranenko]{Lacey:2013is}
Lacey, R.A.; Gu, Y.; Gong, X.; Reynolds, D.; Ajitanand, N.N.; Alexander, J.M.;
  Mwai, A.; Taranenko, A.
\newblock {Is anisotropic flow really acoustic?} {\bf 2013}.
\newblock  \href{http://xxx.lanl.gov/abs/1301.0165}{{\normalfont [1301.0165]}}.

\bibitem[Bhalerao and Ollitrault(2006)]{Olli}
Bhalerao, R.S.; Ollitrault, J.Y.
\newblock {Eccentricity fluctuations and elliptic flow at RHIC}.
\newblock {\em Phys. Lett.} {\bf 2006}, {\em B641},~260--264,
  \href{http://xxx.lanl.gov/abs/nucl-th/0607009}{{\normalfont
  [arXiv:nucl-th/nucl-th/0607009]}}.

\bibitem[Staig and Shuryak(2011)]{Staig:2011wj}
Staig, P.; Shuryak, E.
\newblock {The Fate of the Initial State Fluctuations in Heavy Ion Collisions.
  III The Second Act of Hydrodynamics}.
\newblock {\em Phys. Rev.} {\bf 2011}, {\em C84},~044912,
  \href{http://xxx.lanl.gov/abs/1105.0676}{{\normalfont
  [arXiv:nucl-th/1105.0676]}}.

\bibitem[Rose \em{et~al.}(2014)Rose, Paquet, Denicol, Luzum, Schenke, Jeon, and
  Gale]{Rose:2014fba}
Rose, J.B.; Paquet, J.F.; Denicol, G.S.; Luzum, M.; Schenke, B.; Jeon, S.;
  Gale, C.
\newblock {Extracting the bulk viscosity of the quark–gluon plasma}.
\newblock {\em Nucl. Phys.} {\bf 2014}, {\em A931},~926--930,
  \href{http://xxx.lanl.gov/abs/1408.0024}{{\normalfont
  [arXiv:nucl-th/1408.0024]}}.

\bibitem[Jia(2011)]{ATLAS_corr}
Jia, J.
\newblock {Measurement of elliptic and higher order flow from ATLAS experiment
  at the LHC}.
\newblock {\em J. Phys.} {\bf 2011}, {\em G38},~124012,
  \href{http://xxx.lanl.gov/abs/1107.1468}{{\normalfont
  [arXiv:nucl-ex/1107.1468]}}.

\bibitem[Ade \em{et~al.}(2014)Ade et~al.]{Ade:2013kta}
Ade, P.A.R.; others.
\newblock {Planck 2013 results. XV. CMB power spectra and likelihood}.
\newblock {\em Astron. Astrophys.} {\bf 2014}, {\em 571},~A15,
  \href{http://xxx.lanl.gov/abs/1303.5075}{{\normalfont
  [arXiv:astro-ph.CO/1303.5075]}}.

\bibitem[Grosse-Oetringhaus(2014)]{ALICE_qm20141}
Grosse-Oetringhaus, J.F.
\newblock {Overview of ALICE Results at Quark Matter 2014}.
\newblock {\em Nucl. Phys.} {\bf 2014}, {\em A931},~22--31,
  \href{http://xxx.lanl.gov/abs/1408.0414}{{\normalfont
  [arXiv:nucl-ex/1408.0414]}}.

\bibitem[Goldhaber \em{et~al.}(1960)Goldhaber, Goldhaber, Lee, and
  Pais]{Goldhaber:1960sf}
Goldhaber, G.; Goldhaber, S.; Lee, W.Y.; Pais, A.
\newblock {Influence of Bose-Einstein statistics on the anti-proton proton
  annihilation process}.
\newblock {\em Phys. Rev.} {\bf 1960}, {\em 120},~300--312.

\bibitem[Kopylov and Podgoretsky(1974)]{Kopylov:1973qq}
Kopylov, G.I.; Podgoretsky, M.I.
\newblock {Multiple production and interference of particles emitted by moving
  sources}.
\newblock {\em Sov. J. Nucl. Phys.} {\bf 1974}, {\em 18},~336--341.
\newblock [Yad. Fiz.18,656(1973)].

\bibitem[Shuryak(1973)]{Shuryak:1974am}
Shuryak, E.V.
\newblock {Correlation of identical pions in multiple production reactions}.
\newblock {\em Yad. Fiz.} {\bf 1973}, {\em 18},~1302--1308.

\bibitem[Pratt(2009)]{Pratt:2008qv}
Pratt, S.
\newblock {Resolving the HBT Puzzle in Relativistic Heavy Ion Collision}.
\newblock {\em Phys. Rev. Lett.} {\bf 2009}, {\em 102},~232301,
  \href{http://xxx.lanl.gov/abs/0811.3363}{{\normalfont
  [arXiv:nucl-th/0811.3363]}}.

\bibitem[Policastro \em{et~al.}(2001)Policastro, Son, and
  Starinets]{Policastro:2001yc}
Policastro, G.; Son, D.T.; Starinets, A.O.
\newblock {The Shear viscosity of strongly coupled N=4 supersymmetric
  Yang-Mills plasma}.
\newblock {\em Phys. Rev. Lett.} {\bf 2001}, {\em 87},~081601,
  \href{http://xxx.lanl.gov/abs/hep-th/0104066}{{\normalfont
  [arXiv:hep-th/hep-th/0104066]}}.

\bibitem[Prakash \em{et~al.}(1993)Prakash, Prakash, Venugopalan, and
  Welke]{Prakash:1993bt}
Prakash, M.; Prakash, M.; Venugopalan, R.; Welke, G.
\newblock {Nonequilibrium properties of hadronic mixtures}.
\newblock {\em Phys. Rept.} {\bf 1993}, {\em 227},~321--366.

\bibitem[Gelman \em{et~al.}(2006)Gelman, Shuryak, and Zahed]{Gelman:2006xw}
Gelman, B.A.; Shuryak, E.V.; Zahed, I.
\newblock {Classical strongly coupled QGP. I. The Model and molecular dynamics
  simulations}.
\newblock {\em Phys. Rev.} {\bf 2006}, {\em C74},~044908,
  \href{http://xxx.lanl.gov/abs/nucl-th/0601029}{{\normalfont
  [arXiv:nucl-th/nucl-th/0601029]}}.

\bibitem[Nakamura and Sakai(2005)]{Nakamura:2004sy}
Nakamura, A.; Sakai, S.
\newblock {Transport coefficients of gluon plasma}.
\newblock {\em Phys. Rev. Lett.} {\bf 2005}, {\em 94},~072305,
  \href{http://xxx.lanl.gov/abs/hep-lat/0406009}{{\normalfont
  [arXiv:hep-lat/hep-lat/0406009]}}.

\bibitem[Ratti and Shuryak(2009)]{Ratti:2008jz}
Ratti, C.; Shuryak, E.
\newblock {The Role of monopoles in a Gluon Plasma}.
\newblock {\em Phys. Rev.} {\bf 2009}, {\em D80},~034004,
  \href{http://xxx.lanl.gov/abs/0811.4174}{{\normalfont
  [arXiv:hep-ph/0811.4174]}}.

\bibitem[Schafer and Shuryak(1998)]{Schafer:1996wv}
Schafer, T.; Shuryak, E.V.
\newblock {Instantons in QCD}.
\newblock {\em Rev. Mod. Phys.} {\bf 1998}, {\em 70},~323--426,
  \href{http://xxx.lanl.gov/abs/hep-ph/9610451}{{\normalfont
  [arXiv:hep-ph/hep-ph/9610451]}}.

\bibitem[Lee and Lu(1998)]{Lee:1998bb}
Lee, K.M.; Lu, C.h.
\newblock {SU(2) calorons and magnetic monopoles}.
\newblock {\em Phys. Rev.} {\bf 1998}, {\em D58},~025011,
  \href{http://xxx.lanl.gov/abs/hep-th/9802108}{{\normalfont
  [arXiv:hep-th/hep-th/9802108]}}.

\bibitem[Kraan and van Baal(1998)]{Kraan:1998sn}
Kraan, T.C.; van Baal, P.
\newblock {Monopole constituents inside SU(n) calorons}.
\newblock {\em Phys. Lett.} {\bf 1998}, {\em B435},~389--395,
  \href{http://xxx.lanl.gov/abs/hep-th/9806034}{{\normalfont
  [arXiv:hep-th/hep-th/9806034]}}.

\bibitem[Liu \em{et~al.}(2015)Liu, Shuryak, and Zahed]{Liu:2015jsa}
Liu, Y.; Shuryak, E.; Zahed, I.
\newblock {Light quarks in the screened dyon-antidyon Coulomb liquid model.
  II.}
\newblock {\em Phys. Rev.} {\bf 2015}, {\em D92},~085007,
  \href{http://xxx.lanl.gov/abs/1503.09148}{{\normalfont
  [arXiv:hep-ph/1503.09148]}}.

\bibitem[Larsen and Shuryak(2016)]{Larsen:2015tso}
Larsen, R.; Shuryak, E.
\newblock {Instanton-dyon Ensemble with two Dynamical Quarks: the Chiral
  Symmetry Breaking}.
\newblock {\em Phys. Rev.} {\bf 2016}, {\em D93},~054029,
  \href{http://xxx.lanl.gov/abs/1511.02237}{{\normalfont
  [arXiv:hep-ph/1511.02237]}}.

\bibitem[Shuryak(2016)]{Shuryak:2016vow}
Shuryak, E.
\newblock {Recent progress in understanding deconfinement and chiral
  restoration phase transitions} {\bf 2016}.
\newblock  \href{http://xxx.lanl.gov/abs/1610.08789}{{\normalfont
  [arXiv:nucl-th/1610.08789]}}.

\bibitem[Seiberg and Witten(1994)]{Seiberg:1994rs}
Seiberg, N.; Witten, E.
\newblock {Electric - magnetic duality, monopole condensation, and confinement
  in N=2 supersymmetric Yang-Mills theory}.
\newblock {\em Nucl. Phys.} {\bf 1994}, {\em B426},~19--52,
  \href{http://xxx.lanl.gov/abs/hep-th/9407087}{{\normalfont
  [arXiv:hep-th/hep-th/9407087]}}.
\newblock [Erratum: Nucl. Phys.B430,485(1994)].

\bibitem[Baker \em{et~al.}(1997)Baker, Ball, and Zachariasen]{Baker:1997bg}
Baker, M.; Ball, J.S.; Zachariasen, F.
\newblock {Comparison of lattice and dual QCD results for heavy quark
  potentials}.
\newblock {\em Phys. Rev.} {\bf 1997}, {\em D56},~4400--4403,
  \href{http://xxx.lanl.gov/abs/hep-ph/9705207}{{\normalfont
  [arXiv:hep-ph/hep-ph/9705207]}}.

\bibitem[Maldacena(1999)]{Maldacena:1997re}
Maldacena, J.M.
\newblock {The Large N limit of superconformal field theories and
  supergravity}.
\newblock {\em Int. J. Theor. Phys.} {\bf 1999}, {\em 38},~1113--1133,
  \href{http://xxx.lanl.gov/abs/hep-th/9711200}{{\normalfont
  [arXiv:hep-th/hep-th/9711200]}}.
\newblock [Adv. Theor. Math. Phys.2,231(1998)].

\bibitem[Dirac(1931)]{Dirac:1931kp}
Dirac, P.A.M.
\newblock {Quantized Singularities in the Electromagnetic Field}.
\newblock {\em Proc. Roy. Soc. Lond.} {\bf 1931}, {\em A133},~60--72.

\bibitem[Schwinger \em{et~al.}(1976)Schwinger, Milton, Tsai, DeRaad, and
  Clark]{Schwinger:1976fr}
Schwinger, J.S.; Milton, K.A.; Tsai, W.y.; DeRaad, Jr., L.L.; Clark, D.C.
\newblock {Nonrelativistic Dyon-Dyon Scattering}.
\newblock {\em Annals Phys.} {\bf 1976}, {\em 101},~451.

\bibitem[Boulware \em{et~al.}(1976)Boulware, Brown, Cahn, Ellis, and
  Lee]{Boulware:1976tv}
Boulware, D.G.; Brown, L.S.; Cahn, R.N.; Ellis, S.D.; Lee, C.k.
\newblock {Scattering on Magnetic Charge}.
\newblock {\em Phys. Rev.} {\bf 1976}, {\em D14},~2708.

\bibitem['t~Hooft(1974)]{'tHooft:1974qc}
't~Hooft, G.
\newblock {Magnetic Monopoles in Unified Gauge Theories}.
\newblock {\em Nucl. Phys.} {\bf 1974}, {\em B79},~276--284.

\bibitem[Polyakov(1974)]{Polyakov:1974ek}
Polyakov, A.M.
\newblock {Particle Spectrum in the Quantum Field Theory}.
\newblock {\em JETP Lett.} {\bf 1974}, {\em 20},~194--195.
\newblock [Pisma Zh. Eksp. Teor. Fiz.20,430(1974)].

\bibitem[Mandelstam(1976)]{Mandelstam:1974pi}
Mandelstam, S.
\newblock {Vortices and Quark Confinement in Nonabelian Gauge Theories}.
\newblock {\em Phys. Rept.} {\bf 1976}, {\em 23},~245--249.

\bibitem['t~Hooft(1978)]{'tHooft:1977hy}
't~Hooft, G.
\newblock {On the Phase Transition Towards Permanent Quark Confinement}.
\newblock {\em Nucl. Phys.} {\bf 1978}, {\em B138},~1--25.

\bibitem[Liao and Shuryak(2007)]{Liao:2006ry}
Liao, J.; Shuryak, E.
\newblock {Strongly coupled plasma with electric and magnetic charges}.
\newblock {\em Phys. Rev.} {\bf 2007}, {\em C75},~054907,
  \href{http://xxx.lanl.gov/abs/hep-ph/0611131}{{\normalfont
  [arXiv:hep-ph/hep-ph/0611131]}}.

\bibitem[Hidaka and Pisarski(2008)]{Hidaka:2008dr}
Hidaka, Y.; Pisarski, R.D.
\newblock {Suppression of the Shear Viscosity in a ''semi'' Quark Gluon
  Plasma}.
\newblock {\em Phys. Rev.} {\bf 2008}, {\em D78},~071501,
  \href{http://xxx.lanl.gov/abs/0803.0453}{{\normalfont
  [arXiv:hep-ph/0803.0453]}}.

\bibitem[D'Alessandro \em{et~al.}(2010)D'Alessandro, D'Elia, and
  Shuryak]{D'Alessandro:2010xg}
D'Alessandro, A.; D'Elia, M.; Shuryak, E.V.
\newblock {Thermal Monopole Condensation and Confinement in finite temperature
  Yang-Mills Theories}.
\newblock {\em Phys. Rev.} {\bf 2010}, {\em D81},~094501,
  \href{http://xxx.lanl.gov/abs/1002.4161}{{\normalfont
  [arXiv:hep-lat/1002.4161]}}.

\bibitem[Liao and Shuryak(2008)]{Liao:2008jg}
Liao, J.; Shuryak, E.
\newblock {Magnetic Component of Quark-Gluon Plasma is also a Liquid!}
\newblock {\em Phys. Rev. Lett.} {\bf 2008}, {\em 101},~162302,
  \href{http://xxx.lanl.gov/abs/0804.0255}{{\normalfont
  [arXiv:hep-ph/0804.0255]}}.

\bibitem[Hindmarsh \em{et~al.}(2014)Hindmarsh, Huber, Rummukainen, and
  Weir]{Hindmarsh:2013xza}
Hindmarsh, M.; Huber, S.J.; Rummukainen, K.; Weir, D.J.
\newblock {Gravitational waves from the sound of a first order phase
  transition}.
\newblock {\em Phys. Rev. Lett.} {\bf 2014}, {\em 112},~041301,
  \href{http://xxx.lanl.gov/abs/1304.2433}{{\normalfont
  [arXiv:hep-ph/1304.2433]}}.

\bibitem[Shuryak and Staig(2013)]{Shuryak:2013uaa}
Shuryak, E.; Staig, P.
\newblock {Sounds of the QCD phase transition}.
\newblock {\em Phys. Rev.} {\bf 2013}, {\em C88},~064905,
  \href{http://xxx.lanl.gov/abs/1306.2938}{{\normalfont
  [arXiv:nucl-th/1306.2938]}}.

\bibitem[Kalaydzhyan and Shuryak(2015)]{ourGW}
Kalaydzhyan, T.; Shuryak, E.
\newblock {Gravity waves generated by sounds from big bang phase transitions}.
\newblock {\em Phys. Rev.} {\bf 2015}, {\em D91},~083502,
  \href{http://xxx.lanl.gov/abs/1412.5147}{{\normalfont
  [arXiv:hep-ph/1412.5147]}}.

\bibitem[Shuryak(1978)]{Shuryak:1978ij}
Shuryak, E.V.
\newblock {Quark-Gluon Plasma and Hadronic Production of Leptons, Photons and
  Psions}.
\newblock {\em Phys. Lett.} {\bf 1978}, {\em B78},~150.
\newblock [Yad. Fiz.28,796(1978)].

\bibitem[Witten(1984)]{Witten:1984rs}
Witten, E.
\newblock {Cosmic Separation of Phases}.
\newblock {\em Phys. Rev.} {\bf 1984}, {\em D30},~272--285.

\end{thebibliography}


\end{document}